\documentclass[a4paper,11pt]{article}
\pdfoutput=1
\usepackage{jheppub}

\usepackage{latexsym}
\usepackage{bbm}
\usepackage{colortbl}

\title{Blackened BPS: the end of the thermal BMN branch}

\author[a]{Jorge~E.~Santos}
\affiliation[a]{Department of Applied Mathematics and Theoretical Physics,\\
University of Cambridge, Wilberforce Road, Cambridge CB3 0WA, United Kingdom}

\abstract{%
We study the thermodynamics of the BMN matrix model at large $N$ and strong coupling, using its dual black holes in eleven-dimensional supergravity. Within the consistent truncation to two-dimensional $SO(9)$ gauged supergravity, the static $SO(3)\times SO(6)$-invariant black holes solve ordinary differential equations, which we solve to high precision. We formulate holographic renormalisation for this truncation and fix the finite counterterms using a supersymmetric solution. This gives the energy, the entropy and the response to the BMN mass independently, and we verify the first law and the Smarr relation along the branch. The thermal branch ends at a critical value of $\mu/T$, which power-law fits place at $4\pi$ with an estimated uncertainty below $10^{-10}$. The first-order transition inferred in earlier work, where the free energy was reconstructed from the entropy, does not survive, and an independent eleven-dimensional calculation agrees with ours over the range it covers. Instead, the free energy, the entropy and the mass response approach their values in the confined phase continuously at the end of the branch, where the solutions become singular. If this branch remains the dominant one up to its end point, the transition therefore coincides with the singularity. Near the end of the branch, the scalar fields follow a solution of the supersymmetric first-order equations, while the metric carries a nontrivial blackening factor. The horizon develops a neck whose profile is universal, with radii growing as the $2/5$ power of the distance from the pinch. Within the range we resolve, the geometry is not that of the Ricci-flat double cone which governs mergers of vacuum black holes.
}

\newcommand{\be}{\begin{equation}}
\newcommand{\bea}{\begin{eqnarray}}
\newcommand{\eea}{\end{eqnarray}}
\newcommand{\beq}{\begin{equation}}
\newcommand{\ee}{\end{equation}}
\newcommand{\eeq}{\end{equation}}

\def\32{{3 \over 2}}

\def\ba{\begin{eqnarray}}
\def\ea{\end{eqnarray}}

\def\simleq{\;\raise0.3ex\hbox{$<$\kern-0.75em\raise-1.1ex\hbox{$\sim$}}\;}
\def\simgeq{\;\raise0.3ex\hbox{$>$\kern-0.75em\raise-1.1ex\hbox{$\sim$}}\;}

\begin{document}
\maketitle
\flushbottom

\section{Introduction}\label{sec:introduction}
Matrix quantum mechanics provides a setting in which the same system can be described by an explicit Hamiltonian and, in a suitable limit, by classical gravity. The BFSS model is the best-known example, but its commuting matrices have flat directions: they can separate without a corresponding increase in the classical potential. The BMN model replaces this difficulty with a mass deformation that preserves sixteen supercharges and lifts those flat directions \cite{Berenstein:2002jq,Costa:2014wya}. At finite $N$ it has a discrete spectrum and a well-defined thermal ensemble. Its parameters are the 't Hooft coupling $\lambda$, the mass $\mu$ and the temperature $T$, so one can ask how its thermodynamics changes as the mass becomes large compared with the thermal scale.

This is also a useful system in which to make holography quantitative. The matrix Hamiltonian is sufficiently explicit to be studied directly, while at large $N$ and strong coupling the deconfined thermal state has a black-hole description. The absence of flat directions is particularly helpful for numerical studies of the matrix model, and its thermal phase structure has been studied extensively with lattice Monte Carlo methods \cite{Catterall:2010gf,Asano:2018nol,Schaich:2020ubh,Bergner:2021goh,Pateloudis:2022oos,Jha:2024rxz}, including in its bosonic sector \cite{Asano:2020yry,Dhindsa:2022uqn}. Holography for the supersymmetric sector has also been developed using spherical D-branes and supersymmetric localisation \cite{Bobev:2024gqg}. More ambitiously, explicit Hilbert-space truncations and quantum algorithms for simulating matrix-model dynamics have been developed \cite{Gharibyan:2020}, while Maldacena has estimated the quantum resources required to probe a regime with a gravitational black-hole description \cite{Maldacena:2023acv}. The thermodynamic quantities computed here provide concrete observables for such a comparison.

The gravitational calculation is nevertheless not simple. The BMN mass changes the asymptotic eleven-dimensional geometry and turns on a four-form flux. The resulting black hole retains $SO(3)\times SO(6)$ symmetry, but its horizon is deformed. In Ref.~\cite{Costa:2014wya} we constructed this branch by solving the eleven-dimensional field equations as a problem in two variables, starting from the non-extremal D0-brane solution. That calculation suggested a crossing of the black-hole free energy and hence a transition to a confined phase. In that work, however, the free energy was not computed directly: it was reconstructed from the horizon entropy by integrating the first law, using a polynomial fit of the entropy. As we show in Appendix~\ref{app:conformal}, this reconstruction is unstable precisely where the crossing was found. Here we compute the free energy directly from the renormalised boundary data, which allows us to determine whether the proposed crossing survives.

The main simplification comes from the consistent truncation to two-dimensional $SO(9)$ gauged supergravity \cite{Ortiz:2012ib,BossardCiceriInversoKleinschmidt}. In the $SO(3)\times SO(6)$ sector the fields of the reduced theory depend, for static solutions, on a single radial coordinate. The nonlinear uplift reconstructs their full eleven-dimensional angular dependence, so the reduced solutions are also solutions of eleven-dimensional supergravity. This gives us a lower-dimensional gravitational model of BMN holography in which the thermal branch can be resolved much more accurately. It does not imply that every black hole of the eleven-dimensional theory lies in this truncation.

To compare the reduced solutions with the matrix model, we must also fix the boundary sources and the definition of the energy. We formulate holographic renormalisation for time-dependent boundary fields in the bosonic truncation, and then specialise to the stationary BMN family. The supersymmetric solution supplies the condition used to select the finite prescription for the thermal charges. This allows us to compute the free energy and the response to the BMN mass from the asymptotic data, while the entropy is obtained at the horizon.

We find a thermal branch along which $\mu/T$ increases monotonically towards a critical value. Power-law fits near the end of the branch place this value at $4\pi$, with an estimated uncertainty below $10^{-10}$, although we have no analytic argument that fixes it. The free energy remains negative at the value where Ref.~\cite{Costa:2014wya} inferred a transition, and an independent eleven-dimensional calculation in conformal coordinates agrees with the reduced results over the range it presently covers. Provided this branch remains the dominant one up to its end point, there is nevertheless a transition, and it occurs at the end of the branch. As $\mu/T$ approaches its critical value, the free energy, the entropy and the mass response of the black holes tend continuously to their values in the confined phase. The free energy does so as $(4\pi-\mu/T)^{8/5}$, so the transition appears to be continuous rather than first order. It coincides with the point at which the solutions become singular, as was found for mass-deformed M2 branes in Ref.~\cite{Dias:2017stenzel}. Near the end of the branch, the scalar fields follow a solution of the supersymmetric first-order equations with high accuracy, while the metric carries a nontrivial blackening factor. The horizon moves into the singular region of this flow, and it develops a pronounced neck. In the range we can resolve, its geometry is not that of the Ricci-flat double cone which governs topology-changing mergers of vacuum black holes. The global phase structure is not settled by this branch alone: other black-hole branches, which might pre-empt this transition, must also be compared with the confined backgrounds.

We begin by recalling the matrix model and its eleven-dimensional description. We then give the reduction and the boundary prescription, construct the supersymmetric and thermal solutions, and present their thermodynamics. The appendices contain the covariant charge calculation, details of the supersymmetric boundary condition, the integration of the explicit supersymmetric flow, and the comparison with the direct eleven-dimensional construction.

\section{The BMN matrix model}\label{sec:bmnmodel}

We begin by recalling the matrix quantum mechanics whose thermal states we wish to describe. Our conventions follow \cite{Costa:2014wya}. The undeformed theory is the BFSS model, obtained by reducing ten-dimensional $SU(N)$ super Yang--Mills theory to one time dimension. Its field content consists of nine traceless Hermitian matrices $X^I$, with $I=1,\ldots,9$, sixteen fermionic matrices $\Psi^\alpha$ and a gauge field $A_t$. Its action reads
\begin{equation}
 S_{\rm BFSS}=\frac{N}{2\lambda}\int {\rm d}t\,{\rm Tr}\left[
 (D_tX^I)^2+\Psi^\alpha D_t\Psi^\alpha
 +\frac{1}{2}[X^I,X^J]^2
 +i\Psi^\alpha\gamma^I_{\alpha\beta}[\Psi^\beta,X^I]
 \right],
 \label{eq:BFSSaction}
\end{equation}
where $D_t=\partial_t-i[A_t,\,\cdot\,]$, the $\gamma^I$ are $SO(9)$ gamma matrices, and repeated indices $I,J=1,\ldots,9$ and $\alpha,\beta=1,\ldots,16$ are summed. The 't Hooft coupling $\lambda$ has dimensions of energy cubed, so that the dimensionless temperature of the undeformed theory is $T/\lambda^{1/3}$.

The BMN model is a one-parameter deformation of this theory which preserves its sixteen supercharges \cite{Berenstein:2002jq}. In the conventions of \cite{Costa:2014wya}, its action reads
\begin{align}
 S_{\rm BMN} = S_{\rm BFSS} -\frac{N}{2\lambda}\int {\rm d}t\,{\rm Tr}\,\bigg[
 &\frac{\mu^2}{3^2}(X^i)^2+\frac{\mu^2}{6^2}(X^a)^2
 +\frac{\mu}{4}\Psi^\alpha(\gamma^{123})_{\alpha\beta}\Psi^\beta
 \nonumber\\
 &\hspace{2cm}+\frac{2i\mu}{3}\epsilon_{ijk}X^iX^jX^k\bigg],
 \label{eq:BMNaction}
\end{align}
where $i,j,k=1,2,3$, $a=4,\ldots,9$, and $\gamma^{123}=\gamma^{[1}\gamma^2\gamma^{3]}$. We take $\mu>0$. The different masses of the two sets of matrices break the $SO(9)$ symmetry of the BFSS model to $SO(3)\times SO(6)$. Notice that the bosonic masses, the fermion mass and the cubic interaction are all fixed by the same parameter. This fact will be important when identifying the boundary conditions in the gravitational description: varying the BMN mass changes these couplings together.

The mass deformation lifts the commuting flat directions of the BFSS model. Its classical supersymmetric vacua are
\begin{equation}
 X^a=0,\qquad X^i=\frac{\mu}{3}J^i,\qquad [J^i,J^j]=i\epsilon_{ijk}J^k,
 \label{eq:BMNvacua}
\end{equation}
where the $J^i$ form an $N$-dimensional representation of $SU(2)$. Decomposing this representation into irreducible blocks shows that the vacua are labelled by partitions of $N$. The trivial vacuum has $J^i=0$. As in \cite{Costa:2014wya}, we will study the thermal black hole branch obtained by continuously deforming the non-extremal D0-brane solution. We will not attempt to describe the thermal phases associated with every choice of vacuum.

At finite temperature, the bosons are periodic and the fermions antiperiodic around the Euclidean time circle, whose circumference is $1/T$. These thermal boundary conditions break supersymmetry, although the relations between the couplings in \eqref{eq:BMNaction} remain unchanged. At fixed $N$, the theory depends on two dimensionless ratios, which we can take to be $\lambda/\mu^3$ and $T/\mu$. We are interested in the planar, strongly coupled regime, with both $T$ and $\mu$ small compared with $\lambda^{1/3}$ and with their ratio held fixed. It is convenient to parametrise the thermal solutions by
\begin{equation}
 \widehat\mu\equiv\frac{\mu}{T}.
 \label{eq:QFTmuhat}
\end{equation}
Small $\widehat\mu$ corresponds to a small mass deformation on the thermal scale. Increasing $\widehat\mu$ at fixed $\mu$ amounts to lowering the temperature. In particular, high temperature relative to $\mu$ need not imply weak coupling: one can have $\mu\ll T\ll\lambda^{1/3}$.

Our aim is to determine the thermodynamics of this branch as a function of $\widehat\mu$. The quantities of interest are its energy $E$, entropy $S$ and free energy $F=E-TS$, together with the response to the mass deformation. At fixed $T$, $N$ and $\lambda$, this response is
\begin{equation}
 \Pi\equiv\left(\frac{\partial F}{\partial\mu}\right)_{T,N,\lambda}
 =\left\langle\frac{\partial H_{\rm BMN}}{\partial\mu}\right\rangle.
 \label{eq:QFTmassresponse}
\end{equation}
Thus, $\Pi$ probes the combination of operators obtained by differentiating the full deformation in \eqref{eq:BMNaction}. The gravitational calculation must identify this combination, as well as the free energy, with the same choice of sources.

\section{The eleven-dimensional gravitational description}\label{sec:gravitydual}

We use mostly-plus Lorentzian signature, $(-,+,\ldots,+)$, throughout. Our curvature convention is
\begin{equation}
 R^a{}_{bcd}=\partial_c\Gamma^a_{db}-\partial_d\Gamma^a_{cb}
 +\Gamma^a_{ce}\Gamma^e_{db}-\Gamma^a_{de}\Gamma^e_{cb},
 \qquad R_{bd}=R^a{}_{bad}.
 \label{eq:curvatureconvention}
\end{equation}
Let us now recall the gravitational problem described in \cite{Costa:2014wya}. At zero mass deformation, the relevant thermal solution is the near-horizon geometry of non-extremal D0-branes. Its uplift to eleven dimensions is purely geometric: the D0-brane charge becomes momentum along the M-theory circle, and the three-form potential vanishes. The metric can be written as
\begin{equation}
 {\rm d}s_{11}^2=\frac{{\rm d}u^2}{h(u)}+u^2{\rm d}\Omega_8^2 +\frac{L^7}{u^7}{\rm d}\psi^2 +h(u)\,{\rm d}t\left(2{\rm d}\psi-\frac{u_0^7}{L^7}{\rm d}t\right), \qquad h(u)=1-\left(\frac{u_0}{u}\right)^7\,,
 \label{eq:D0eleven}
\end{equation}
where, throughout, ${\rm d}\Omega_n^2$ denotes the metric on a round unit-radius $n$-sphere. Here $\psi$ parametrises the M-theory circle, $L$ is the length scale fixed by the D0-brane charge, and $u=u_0$ is the horizon. We use $u$ for the eleven-dimensional radial coordinate to distinguish it from the radial coordinates of the reduced theory below. The horizon has topology $S^8\times S^1$, and its temperature is
\begin{equation}
 T=\frac{7}{4\pi}\frac{u_0^{5/2}}{L^{7/2}}.
 \label{eq:D0eleventemperature}
\end{equation}
The $SO(9)$ symmetry acts on the round $S^8$.

Turning on the BMN deformation requires a nonzero four-form field strength $F_{(4)}={\rm d}C_{(3)}$. The mass parameter is fixed by the asymptotic plane-wave data. With the same signature as \eqref{eq:D0eleven}, a convenient form of the plane wave is
\begin{subequations}
\begin{align}
 {\rm d}s_{\rm pw}^2
 &=\sum_{i=1}^3({\rm d}x^i)^2+\sum_{a=4}^9({\rm d}x^a)^2
 +2{\rm d}t\,{\rm d}\psi
 -\left[\frac{\mu^2}{9}\sum_{i=1}^3(x^i)^2
 +\frac{\mu^2}{36}\sum_{a=4}^9(x^a)^2\right]{\rm d}t^2,
 \label{eq:BMNplanewavemetric}\\
 F_{(4)}&=\mu\,{\rm d}t\wedge{\rm d}x^1\wedge{\rm d}x^2\wedge{\rm d}x^3.
 \label{eq:BMNplanewaveflux}
\end{align}
\label{eq:BMNplanewave}
\end{subequations}
The orientation has been chosen to give the sign in \eqref{eq:BMNplanewaveflux}. These expressions exhibit the same $SO(3)\times SO(6)$ splitting as the matrix-model action. They specify the deformation at infinity; the interior geometry must be determined by solving the field equations with a regular horizon. In particular, the thermal solution carries both the circle momentum associated with $N$ and the asymptotic flux associated with $\mu$.

The symmetry reduction of the eight-sphere is easily described by writing its unit-radius metric as
\begin{equation}
 {\rm d}\Omega_8^2=\frac{{\rm d}\zeta^2}{1-\zeta^2}
 +\zeta^2{\rm d}\Omega_2^2+(1-\zeta^2){\rm d}\Omega_5^2,
 \qquad 0\leq\zeta\leq1.
 \label{eq:roundS8split}
\end{equation}
The $S^2$ collapses at $\zeta=0$, while the $S^5$ collapses at $\zeta=1$. For a stationary geometry invariant under translations along the M-theory circle, imposing $SO(3)\times SO(6)$ symmetry still leaves dependence on a radial coordinate and on $\zeta$. The direct construction of \cite{Costa:2014wya} therefore involved ten unknown functions of two variables. Regularity was imposed at the horizon and at both symmetry axes, together with the BMN source conditions at infinity, and the resulting Einstein--DeTurck and form-field equations were solved numerically.

That construction provides the comparison for the solutions studied here. There is one convention which should be kept in mind: the parameter denoted by $\widehat\mu$ in \cite{Costa:2014wya} is related to ours by
\begin{equation}
 \widehat\mu_{\rm CGPS}=\frac{7}{12\pi}\frac{\mu}{T}
 =\frac{7}{12\pi}\widehat\mu.
 \label{eq:CGPSmassconversion}
\end{equation}
We will use \eqref{eq:QFTmuhat} throughout. As we discuss in Appendix~\ref{app:conformal}, however, the free-energy curve plotted in Ref.~\cite{Costa:2014wya} is reproduced at small mass only if its horizontal axis is rescaled with respect to \eqref{eq:CGPSmassconversion}, and we compare with those data in that way.

The direct eleven-dimensional problem becomes substantially simpler within the consistent truncation to two-dimensional $SO(9)$ gauged supergravity \cite{Ortiz:2012ib,BossardCiceriInversoKleinschmidt}. In the sector relevant here, the independent fields are a two-dimensional metric, a dilaton and two scalars. Their full dependence on the internal coordinates is supplied by the nonlinear uplift formulae. Consequently, finding static solutions requires solving ordinary differential equations in a single radial coordinate. The angular deformation of the eleven-dimensional geometry is retained by the uplift, including its nonlinear dependence on the two-dimensional fields.

This reduction allows us to concentrate the numerical resolution on the radial problem and to follow the thermal branch as the mass deformation is increased. Consistency guarantees that every solution of the reduced equations uplifts to a solution of eleven-dimensional supergravity. It does not imply that the truncation contains every solution with the same global symmetries. In the following section we present the eleven-dimensional Ansatz and the reduced action, which together specify the class of geometries considered in this work.

\section{Reduction to two dimensions}\label{sec:reduction}
\subsection{Eleven-dimensional Ansatz}\label{sec:uplift}
The eleven-dimensional Ansatz follows from the non-linear uplift of two-dimensional $SO(9)$ gauged supergravity constructed in \cite{BossardCiceriInversoKleinschmidt}. In the $SO(3)\times SO(6)$ invariant sector considered here, it is specified by a two-dimensional metric $\mathrm{d}s_2^2$, a dilaton $\rho$, a scalar $v$ and a pseudoscalar $\phi$. These fields depend only on the two-dimensional coordinates. The dependence on the internal coordinates is fixed by the Ansatz below.

We take the two-dimensional volume form to be ${\rm vol}_2=\sqrt{-g}\,{\rm d}t\wedge{\rm d}r$.

Using the decomposition of the round $S^8$ in \eqref{eq:roundS8split}, we write its embedding coordinates as
\begin{equation}
 Y^a=\zeta\,Y^a_2,\qquad Y^{\hat a}=\sqrt{1-\zeta^2}\,Y^{\hat a}_5,  \qquad  a=1,2,3,\qquad \hat a=4,\ldots,9.
 \label{eq:S8coordinates}
\end{equation}
Here $Y_2^a$ and $Y_5^{\hat a}$ are embedding coordinates on unit $S^2$ and $S^5$, respectively. We define
\begin{equation}
 \Delta  \equiv  \zeta^2e^{-2v}  +(1-\zeta^2)e^{v},  \qquad  {\cal H}  \equiv  1+\frac{4\zeta^2e^{4v}\phi^2}  {\rho^{2/3}\Delta}.
 \label{eq:DeltaF}
\end{equation}

The eleven-dimensional metric is
\begin{equation}
 {\rm d}s_{11}^2=g^2\Delta{\cal H} ^{1/3}\,{\rm d}s_2^2 +\rho^{4/9}{\cal H} ^{1/3} \bigg[  e^{v}\Delta\frac{{\rm d}\zeta^2}{1-\zeta^2}  +\frac{e^{2v}}{{\cal H} }\,    \zeta^2{\rm d}\Omega_2^2  +e^{-v}(1-\zeta^2){\rm d}\Omega_5^2
 \bigg] +\frac{\rho^{-14/9}}{\Delta{\cal H}^{2/3}}\,  {\cal D}\psi^2 ,
 \label{eq:11dmetric}
\end{equation}
where $\psi$ parametrises the M-theory circle and
\begin{equation}
 {\cal D}\psi  =  {\rm d}\psi  +\zeta^2\omega_3  +(1-\zeta^2)\omega_6 .
 \label{eq:Dpsi}
\end{equation}
The two one-forms $\omega_3$ and $\omega_6$ live on the two-dimensional spacetime and are determined by
\begin{subequations}
\label{eq:omegauplift}
\begin{align}
 {\rm d}\omega_3  &=  -g^2\rho^{5/9}  \left( e^{-4v} +6e^{-v} +4e^{2v}\rho^{-2/3}\phi^2  \right)  {\rm vol}_2 , 
 \\
  {\rm d}\omega_6 &=  -g^2\rho^{5/9}  \left(  4e^{2v}+3e^{-v}  \right)  {\rm vol}_2 .
\end{align}
\end{subequations}

The three-form potential is
\begin{equation}
 C_{(3)}  =  \left(  \frac{2\rho^{-2/3}e^{4v}\phi}  {\Delta{\cal H} }\,  {\cal D}\psi  -{\cal A}_1  \right)  \wedge  \zeta^3\,{\rm Vol}_{S^2},
 \label{eq:11dthreeform}
\end{equation}
where ${\rm Vol}_{S^2}$ denotes the volume form on a round unit-radius two-sphere, and the one-form ${\cal A}_1$ on the two-dimensional spacetime satisfies
\begin{equation}
 {\rm d}{\cal A}_1  =  -2g^2\rho^{-1/9}e^{2v}\phi\,  {\rm vol}_2\,.
 \label{eq:A1uplift}
\end{equation}
We choose $y_1=\cos\theta_1$ on the unit $S^2$ and orient it by ${\rm Vol}_{S^2}={\rm d}y_1\wedge{\rm d}\varphi_1$. Our convention for ${\cal A}_1$ and this orientation give the relative minus sign in \eqref{eq:11dthreeform} and the sign of the plane-wave flux in \eqref{eq:BMNplanewaveflux}. For $\rho>0$ and finite $v$ and $\phi$, the internal eight-dimensional space is a squashed $S^8$, with the $S^2$ and $S^5$ collapsing smoothly at $\zeta=0$ and $\zeta=1$, respectively. Equations \eqref{eq:11dmetric}--\eqref{eq:A1uplift} solve the eleven-dimensional field equations whenever the two-dimensional fields obey the equations given below.

The uplift also fixes the normalisation of the BMN mass. Matching the undeformed solution to \eqref{eq:D0eleven} gives $g^2L^7=1$ and identifies the eleven-dimensional radius as $u=gr$. In the static radial gauge introduced below, let $\phi=m_\phi/r+\ldots$. The leading time component of \eqref{eq:11dthreeform} receives contributions from both ${\cal D}\psi$ and ${\cal A}_1$. They are respectively $2g^5m_\phi r^3$ and $-2g^5m_\phi r^3/3$, so that the resulting plane-wave four-form has the normalisation of \eqref{eq:BMNplanewaveflux} when
\begin{equation}
 \mu=4g^2m_\phi .
 \label{eq:massuplift}
\end{equation}
We use $\mu$ for this physical BMN mass throughout; $m_\phi$ denotes only the leading coefficient of the reduced scalar.

\subsection{Two-dimensional action and equations of motion}
To obtain the two-dimensional action, we restrict the $U(1)^4$ truncation of \cite{Anabalon:2013zka} to the fields appearing in the Ansatz. We set the gauge fields $A^{(a)}$ and auxiliary fields $y^{(a)}$ to zero, for $a=1,2,3,4$, and take
\begin{equation}
v_1=v\,,\quad v_2=v_3=v_4=-\frac{v}{2}\,, \quad \phi^{(1)}=\phi\quad\text{and}\quad \phi^{(2)}=\phi^{(3)}=\phi^{(4)}=0\,.
\end{equation}

The maximal $SO(9)$ gauged theory contains scalars transforming in the ${\bf 44}\oplus{\bf 84}$ of $SO(9)$. Restricting to the $SO(3)\times SO(6)$ invariant sector leaves one singlet in each representation. In our truncation, $v$ is the singlet contained in the ${\bf 44}$, whereas $\phi$ is the singlet contained in the ${\bf 84}$; see also \cite{Bovon:2025matrix}. The corresponding matrix-model operators have the schematic form
\begin{equation}
 {\cal O}_{44}  \sim  {\rm Tr}\!\left( X^{(I}X^{J)} -\frac{1}{9}\delta^{IJ}X^KX^K \right), \qquad {\cal O}_{84} \sim {\rm Tr}\!\left(X^{[I}X^JX^{K]}\right) +\text{fermions}.
 \label{eq:matrixoperators}
\end{equation}

The scalar matrix of the $U(1)^4$ truncation is
\begin{equation}
 {\cal V}=\exp\left(v_a h^a\right).
\end{equation}
Using the scalar restriction above, and grouping the directions $(1,2,9)$ together, one finds
\begin{equation}
 {\cal M}\equiv  \left({\cal V}{\cal V}^{T}\right)^{-1}
 = e^{-2v}\,\mathbbm{1}_3 \oplus e^v\,\mathbbm{1}_6 .
 \label{eq:scalarMatrixUplift}
\end{equation}
The scalar matrix convention entering the uplift of
\cite{BossardCiceriInversoKleinschmidt} is the inverse of the convention used in \eqref{eq:scalarMatrixUplift}. Thus, in the notation of that reference,
\begin{equation}
 {\cal M}^{-1}
 = e^{2v}\,\mathbbm{1}_3 \oplus e^{-v}\,\mathbbm{1}_6
 = e^{-2\varphi}\,\mathbbm{1}_3\oplus e^{\varphi}\,\mathbbm{1}_6 ,
\end{equation}
which gives
\begin{equation}
 \varphi=-v\,.
\end{equation}The only non-vanishing axion is $\phi^{129}=\phi$. The Ansatz in section~\ref{sec:uplift} uses these fields, with the following dictionary to the variables of \cite{BossardCiceriInversoKleinschmidt}:
\begin{equation}
 \varrho=\rho,\qquad  \varphi=-v,\qquad  a=2\phi .
 \label{eq:upliftDictionary}
\end{equation}
With the curvature convention \eqref{eq:curvatureconvention}, the resulting action is
\begin{subequations}
\begin{align}
S&=\frac{1}{2\kappa^2_2}\int_{\mathcal{M}}\sqrt{-g}\left(\frac{R}{4}\rho-\frac{9}{8}\rho \nabla_a v\nabla^a v-\frac{e^{6v}}{2}\rho^{1/3}\nabla_a \phi \nabla^a\phi-V\right)\mathrm{d}^2x\nonumber
\\
&=\frac{1}{2\kappa^2_2}\int_{\mathcal{M}}\left(\frac{\rho R}{4}\star 1-\frac{9}{8}\rho\,\mathrm{d}v\wedge \star\mathrm{d}v-\frac{e^{6v}}{2}\rho^{1/3}\,\mathrm{d}\phi\wedge \star\mathrm{d}\phi-V\star 1\right)\,.
\label{eq:actionbmn}
\end{align}
with
\begin{equation}
V=-\frac{3 e^{-4 v}}{8 \rho^{1/9}} \,g^2\,\left[\rho ^{2/3} \left(1+12 e^{3 v}+8 e^{6 v}\right)+4 e^{6 v} \phi ^2\right]\,.
\end{equation}
\end{subequations}

The Hodge dual is defined by $\alpha\wedge\star\beta=\langle\alpha,\beta\rangle_g\,{\rm vol}_2$ for forms of equal degree. Thus, $\star 1={\rm vol}_2$, $\star{\rm vol}_2=-1$, and $\star^2=1$ on one-forms.

As a check of the axion normalisation in \eqref{eq:upliftDictionary}, the axion kinetic term of \cite{BossardCiceriInversoKleinschmidt} reduces, in our normalisation and signature, to
\begin{equation}
 -\frac18\,\rho^{1/3}e^{-6\varphi}(\partial a)^2  =  -\frac12\,\rho^{1/3}e^{6v}(\partial\phi)^2,
\end{equation}
where we used $\varphi=-v$ and $a=2\phi$. This agrees precisely with \eqref{eq:actionbmn}.

Varying \eqref{eq:actionbmn} with respect to $\rho$, $v$, $\phi$ and $g_{ab}$ gives, respectively,
\begin{subequations}
\begin{equation}
R-\frac{9}{2}\nabla_a v\nabla^a v-\frac{2}{3}\frac{e^{6v}}{\rho^{2/3}}\nabla_a \phi\nabla^a \phi-4\frac{\partial V}{\partial \rho}=0\,,
\label{eq:rho}
\end{equation}
\begin{equation}
\nabla^a(\rho \nabla_a v)-\frac{4}{3}e^{6 v}\rho^{1/3}\nabla_a \phi \nabla^a \phi-\frac{4}{9}\frac{\partial V}{\partial v}=0\,,
\end{equation}
\begin{equation}
\nabla^a(e^{6v}\rho^{1/3}\nabla_a \phi)-\frac{\partial V}{\partial \phi}=0\,,
\end{equation}
and
\begin{multline}
R_{ab}-\frac{g_{ab}}{2}R-\frac{1}{\rho}\nabla_a\nabla_b \rho+g_{ab}\frac{\Box \rho}{\rho}=
\\
\frac{9}{2}\left(\nabla_av\nabla_b v-\frac{1}{2}g_{ab}\nabla_c v\nabla^c v\right)+\frac{2\,e^{6v}}{\rho^{2/3}}\left(\nabla_a \phi \nabla_b \phi-\frac{1}{2}g_{ab}\nabla_c \phi \nabla^c \phi\right)-\frac{2\,V}{\rho}\,g_{ab}\,,
\label{eq:mog}
\end{multline}
\label{eqs:truncation}
\end{subequations}
where $\Box=\nabla_a\nabla^a$. The boundary terms in the variation are collected in Appendix~\ref{app:wald}.

\subsection{Static Ansatz and asymptotic data}\label{sec:boundarydata}

We now specialise to static solutions. We write the two-dimensional metric as
\begin{equation}
\mathrm{d}s^2 = -e^{2 A(r)}\mathrm{d}t^2+e^{2 Z(r)}\mathrm{d}r^2
\label{eq:metricAnsatz}
\end{equation}
together with $\rho=\rho(r)$, $v=v(r)$ and $\phi=\phi(r)$.

To specify the asymptotic data, we choose a radial gauge adapted to the D0 boundary and introduce a positive reference scale $r_+$. For the thermal solutions of section~\ref{sec:thermal}, this scale will be the horizon radius. Writing $w=r_+/r$, we parametrise the fields as
\begin{subequations}\label{eq:staticparametrisation}
\begin{align}
 e^{2A}&=(gr)^7(1-w^7)q_1(w),
 &e^{2Z}&=\frac{q_1(w)}{1-w^7},\\
 \rho&=(grq_2(w))^{9/2},
 &e^v&=1+w^2q_3(w),\\
 \phi&=g^{3/2}r_+^{3/2}wq_4(w).
\end{align}
\end{subequations}
Near the boundary, $w=0$, the equations admit an expansion of the form
\begin{equation}
q_i(w) = \sum_{j=0}^{+\infty}a^i_jw^{j}\,.
\label{eq:asymptoticqi}
\end{equation}
The D0 asymptotics require $a^1_0=a^2_0=1$. We denote the dimensionless leading coefficient of the mass deformation by $a^4_0=\bar{\mu}$. Solving the equations order by order through $a^i_{10}$ determines the expansion in terms of four further constants, which we choose to be $a^1_7$, $a^3_0$, $a^3_3$ and $a^4_2$; for instance, $a^1_4=-\tfrac{14}{13}(a^3_0)^2$ and $a^2_4=-\tfrac{2}{13}(a^3_0)^2$. Together with $\bar\mu$, the asymptotic data therefore consist of five real constants. No half-integer powers of $w$ arise: the only such mode of the linearised equations about the D0 solution behaves as $w^{-7/2}$ and is excluded by the D0 asymptotics. The remaining scalar source condition will be determined by the variational principle in section~\ref{sec:boundaryconditions}.

The leading term in $\phi$ then gives
\begin{equation}
 \phi=\frac{m_\phi}{r}+\ldots,
 \qquad
 m_\phi=g^{3/2}r_+^{5/2}\bar\mu,
 \qquad
 \mu=4g^{7/2}r_+^{5/2}\bar\mu .
 \label{eq:mudef}
\end{equation}

\section{Holographic renormalisation and thermodynamic observables}\label{sec:renormalisation}

We formulate the radial variational problem for fields with arbitrary time dependence within the $SO(3)\times SO(6)$-invariant bosonic truncation. The counterterms and source definitions are local in the boundary fields. Only after establishing that prescription do we specialise to the stationary BMN solutions and evaluate their thermodynamic observables.

\subsection{The radial Hamilton--Jacobi problem}\label{sec:hamiltonjacobi}

At a timelike radial cutoff, let $\gamma_{tt}$ be the induced metric and $\gamma$ its determinant. The dilatonic Gibbons--Hawking term is
\begin{equation}
 S_{\rm GH}
 =
 \frac{1}{2\kappa_2^2}
 \int_{\partial{\cal M}}{\rm d}t\,\sqrt{-\gamma}\,
 \frac{\rho}{2}K,
 \label{eq:GHterm}
\end{equation}
with $n^a$ the outward spacelike unit normal, $n^an_a=1$, and $K=\gamma^{ab}\nabla_a n_b$. In radial ADM variables,
\begin{equation}
 {\rm d}s^2=\tilde{N}^2{\rm d}r^2+\gamma_{tt}({\rm d}t+\tilde{N}^t{\rm d}r)^2,
 \qquad \gamma_{tt}<0,
 \label{eq:radialADM}
\end{equation}
we have $K_{tt}=(\partial_r\gamma_{tt}-2D_t\tilde{N}_t)/(2\tilde{N})$. The shift $\tilde{N}^t$ enforces the momentum constraint. In the zero-shift gauge, $\tilde{N}=e^Z$, $\gamma_{tt}=-e^{2A}$, and $K=e^{-Z}A'$, now allowing $A$ and $Z$ to depend on both $r$ and $t$.

Substituting the zero-shift form of \eqref{eq:radialADM} into \eqref{eq:actionbmn} and adding \eqref{eq:GHterm} cancels the radial total derivative. After also integrating by parts in time, the action becomes
\begin{equation}
 \begin{aligned}
 S_{\rm red}=\frac{1}{2\kappa_2^2}\int {\rm d}r\,{\rm d}t\,\Big[&e^{A-Z}\Big(\frac12 A'\rho'-\frac98\rho v'^2-\frac12e^{6v}\rho^{1/3}\phi'^2\Big)\\
 &+e^{Z-A}\Big(\frac98\rho\dot v^2-\frac12\dot\rho\dot Z+\frac12e^{6v}\rho^{1/3}\dot\phi^2\Big) -e^{A+Z}V\Big].
 \end{aligned}
 \label{eq:reducedaction}
\end{equation}
Here and throughout this section a dot denotes $\partial_t$; it is unrelated to the proper-radial derivative denoted by a dot in Appendix~\ref{app:killingspinors}. We discard total time derivatives, as appropriate for a periodic thermal circle or fixed endpoint data. The term involving $\dot Z$ arises from the dilaton-curvature coupling and is needed in the lapse constraint.

The radial canonical momenta, with the overall factor of $1/(2\kappa_2^2)$ stripped off, are
\begin{align}
 \pi_A &= \frac{1}{2}e^{A-Z}\rho',
 &
 \pi_\rho &= \frac{1}{2}e^{A-Z}A',
 \nonumber\\
 \pi_v &= -\frac{9}{4}e^{A-Z}\rho\,v',
 &
 \pi_\phi &= -e^{A-Z}e^{6v}\rho^{1/3}\phi'.
 \label{eq:canonicalmomenta}
\end{align}
For general time dependence, variation of the radial lapse gives the local constraint
\begin{multline}
 0=\frac{1}{-\gamma}\left(2\pi_A\pi_\rho-\frac{2}{9\rho}\pi_v^2-\frac12e^{-6v}\rho^{-1/3}\pi_\phi^2\right)+V+\frac98\rho(Dv)^2+\frac12e^{6v}\rho^{1/3}(D\phi)^2+\frac12\Box_\gamma\rho .
 \label{eq:fullradialconstraint}
\end{multline}
Here $D$ and $\Box_\gamma$ are built from the one-dimensional metric $\gamma_{tt}$, and $\pi_A$ is conjugate to $A=\frac12\log(-\gamma_{tt})$. The last term comes from the time derivative of the radial lapse in \eqref{eq:reducedaction}.
The radial shift also imposes
\begin{equation}
 \partial_t\pi_A-\dot A\pi_A
 -\dot\rho\pi_\rho-\dot v\pi_v-\dot\phi\pi_\phi=0.
 \label{eq:radialmomentumconstraint}
\end{equation}
The local covariant Hamilton--Jacobi functionals below satisfy this identity automatically. At zero boundary-derivative order, inserting
\begin{equation}
 S_{\rm HJ}^{(0)}
 =\frac{1}{2\kappa_2^2}\int {\rm d}t\,\sqrt{-\gamma}\,W(\rho,v,\phi)
 \label{eq:HJAnsatz}
\end{equation}
into \eqref{eq:fullradialconstraint} gives
\begin{equation}
 V =\frac{2}{9\rho}\left(\partial_v W\right)^2
 +\frac{1}{2}e^{-6v}\rho^{-1/3}
 \left(\partial_\phi W\right)^2-2W\,\partial_\rho W.
 \label{eq:HJequation}
\end{equation}
Remarkably, the full nonlinear equation \eqref{eq:HJequation} admits the following simple solution:
\begin{equation}
 W = \frac{3g}{4}\rho^{7/9}e^{-2v}\,\Xi, \qquad \Xi^2 = \left(1+2e^{3v}\right)^2 +\frac{4e^{6v}\phi^2}{\rho^{2/3}} . \label{eq:superpotential}
\end{equation}
This local Hamilton--Jacobi branch will be identified with the supersymmetric flow when we specialise to stationary BMN solutions below.

The corresponding zero-derivative counterterm action is
\begin{equation}  S_{\rm ct}^{\rm susy}  =  -\frac{1}{2\kappa_2^2}  \int_{\partial{\cal M}}{\rm d}t\,\sqrt{-\gamma}\,  W.
 \label{eq:susycounterterm}
\end{equation}
The superscript anticipates the identification with the BPS flow in the stationary specialisation below.

The rescaled axion used below is
\begin{equation}
 x\equiv\frac{\phi}{\rho^{1/3}}.
 \label{eq:defx}
\end{equation}
Its square, $\eta\equiv x^2=\phi^2\rho^{-2/3}$, is the variable used for the supersymmetric flows in section~\ref{sec:susy}.

\subsection{Derivative counterterms for time-dependent fields}\label{sec:time-dependent-renormalisation}

For general boundary data we may allow the leading metric and dilaton to depend on time. In a radial gauge with the D0 falloffs, write
\begin{equation}
 e^A=(gr)^{7/2}a(t)[1+o(1)],\qquad
 \rho=(gr)^{9/2}b(t)[1+o(1)],\qquad
 v=O(r^{-2}),\quad x=O(r^{-5/2}),
 \label{eq:time-dependent-falloffs}
\end{equation}
with smooth positive $a$ and $b$. The leading coefficients of $v$ and $x$ may likewise depend on time. The zero-derivative counterterm alone is insufficient when $\dot b\ne0$.

The two-derivative term forced by the leading dilaton sector of \eqref{eq:fullradialconstraint} can be found without assuming stationarity. Set $v=x=0$ for the moment and define $W_0=\frac94g\rho^{7/9}$. Consider
\begin{equation}
 S_{\rm HJ}^{(2)}=\frac{1}{2\kappa_2^2}
 \int {\rm d}t\,\sqrt{-\gamma}\,f(\rho)(D\rho)^2.
 \label{eq:HJtwo-derivative-ansatz}
\end{equation}
The corresponding functional derivatives, with the common normalisation stripped off, are
\begin{equation}
 \frac{\pi_A^{(2)}}{\sqrt{-\gamma}}=-f(D\rho)^2,
 \qquad
 \frac{\pi_\rho^{(2)}}{\sqrt{-\gamma}}
 =-2f\Box_\gamma\rho-f'(D\rho)^2.
 \label{eq:HJtwo-derivative-momenta}
\end{equation}
The part of \eqref{eq:fullradialconstraint} linear in $S_{\rm HJ}^{(2)}$ is therefore
\begin{equation}
 \left(\frac12-4W_0f\right)\Box_\gamma\rho
 -2(W_0f)'(D\rho)^2=0.
 \label{eq:HJtwo-derivative-equation}
\end{equation}
Both tensor structures vanish for $f=1/(8W_0)$. Consequently the divergent two-derivative counterterm is
\begin{equation}
 S_{\rm ct}^{(2)}=-\frac{1}{2\kappa_2^2}
 \int {\rm d}t\,\sqrt{-\gamma}\,
 \frac{(D\rho)^2}{8W_0}
 =-\frac{1}{2\kappa_2^2}
 \int {\rm d}t\,\sqrt{-\gamma}\,
 \frac{\rho^{-7/9}}{18g}(D\rho)^2.
 \label{eq:time-dependent-counterterm}
\end{equation}
It respects the scaling covariance \eqref{eq:dilatonscaling}. With \eqref{eq:time-dependent-falloffs} its density is $O(r^2)$ when $\dot b\ne0$; hence it cannot be omitted for general time-dependent sources. It is zero when the boundary fields are independent of time.

Relative to the $O(r^7)$ density $\sqrt{-\gamma}W_0$, each pair of boundary derivatives in the pure-dilaton Hamilton--Jacobi expansion lowers the weight by five powers of $r$. Four-derivative terms therefore vanish at the cutoff. Derivatives of the scalar sources are likewise radially integrable under \eqref{eq:time-dependent-falloffs}. The two-derivative Hamilton--Jacobi recursion produces no additional finite term linear in $v$ on the analytic D0 branch; the relevant power counting and recursion are given in Appendix~\ref{app:countertermscaling}.

For the bosonic problem with the falloffs \eqref{eq:time-dependent-falloffs}, the local divergent prescription is $S_{\rm ct}^{\rm susy}+S_{\rm ct}^{(2)}$. The second term contributes to the renormalised metric and dilaton momenta through \eqref{eq:HJtwo-derivative-momenta}.

\subsection{Finite counterterms and boundary sources}\label{sec:finitecounterterms}

Expanding \eqref{eq:superpotential} about the D0 solution gives
\begin{align}
 \frac{W}{g\rho^{7/9}}
 ={}& \frac{9}{4} +\frac{9}{4}v^2 +\frac{1}{2}x^2 -\frac{3}{4}v^3 +v x^2 +\frac{9}{16}v^4 +\frac{1}{2}v^2x^2 -\frac{1}{18}x^4 +\ldots .
 \label{eq:superpotentialexpansion}
\end{align}
The mixed term $v x^2$ will play a particularly important role below. In terms of the original fields it gives
\begin{equation}
 W\supset g\rho^{1/9}v\phi^2 .
 \label{eq:finitecounterterm}
\end{equation}
This term gives a finite contribution at the D0 boundary, so cancellation of divergences does not determine its coefficient. To identify the remaining freedom, we restrict to local counterterms without boundary derivatives of the form
\begin{equation}
 U(\rho,v,\phi)=g\rho^{7/9}u(v,x),
 \qquad u(v,x)\ \text{analytic near}\ (v,x)=(0,0).
 \label{eq:countertermclass}
\end{equation}
We require the counterterms to respect the scaling covariance of the classical bulk action, derived in Appendix~\ref{app:countertermscaling}; see Ref.~\cite{Bobev:2026kld} for a general discussion of scale covariance in D$p$-brane holography. This fixes their dilaton homogeneity to that of $W$. Analyticity near the D0 solution is a separate requirement. We also keep the quantisations and the BMN deformation fixed.

For the time-dependent falloffs \eqref{eq:time-dependent-falloffs}, $\sqrt{-\gamma}\,g\rho^{7/9}=O(r^7)$, $v=O(r^{-2})$ and $x=O(r^{-5/2})$ pointwise in time. A monomial in the counterterm density therefore behaves as
\begin{equation}
 \sqrt{-\gamma}\,g\rho^{7/9}v^m x^n
 =O\!\left(r^{7-2m-5n/2}\right).
 \label{eq:countertermweight}
\end{equation}
Terms with $2m+5n/2<7$ are fixed by cancelling divergences for general asymptotic data. Those with $2m+5n/2>7$ vanish in the renormalised action. The only non-negative integer solution of $2m+5n/2=7$ is $(m,n)=(1,2)$. Thus, within \eqref{eq:countertermclass}, the only finite zero-derivative freedom is the coefficient of $vx^2$, even when its leading coefficients depend on time.

The Hamilton--Jacobi equation leaves precisely this coefficient undetermined at the order at which it contributes. Indeed, substituting \eqref{eq:countertermclass} into \eqref{eq:HJequation} gives
\begin{equation}
 -\frac{3}{8}e^{-4v}\left(1+12e^{3v}+8e^{6v}+4e^{6v}x^2\right)
 =\frac{2}{9}u_v^2+\frac12e^{-6v}u_x^2
 -\frac{14}{9}u^2+\frac23xu u_x.
 \label{eq:dimensionlessHJ}
\end{equation}
At the order of a new monomial, the linear operator about the D0 branch is $2v\partial_v+\frac52x\partial_x-7$, which annihilates $vx^2$. In particular, inserting
\begin{equation}
 u=\frac94+\frac94v^2+\frac12x^2-\frac34v^3+c\,vx^2+\ldots
 \label{eq:HJfinitefreedom}
\end{equation}
into \eqref{eq:dimensionlessHJ} leaves $c$ free at this order. An additional boundary condition is therefore needed to select $c=1$, the value in \eqref{eq:superpotentialexpansion}. We impose it after specialising to the BMN family.

We parameterise this remaining freedom by
\begin{equation}
 U_c=W+(c-1)g\rho^{1/9}v\phi^2.
 \label{eq:countertermc}
\end{equation}
The canonical momenta, evaluated for the $W$ branch, define
\begin{align}
 \Pi_v^{\rm ren} &= \pi_v-\sqrt{-\gamma}\,\partial_vW,
 \label{eq:Pivren}\\
 \Pi_\phi^{\rm ren} &= \pi_\phi-\sqrt{-\gamma}\,\partial_\phi W.
 \label{eq:Phiphiren}
\end{align}
The derivative counterterm \eqref{eq:time-dependent-counterterm} has no explicit $v$ or $\phi$ dependence and therefore does not change these two momenta. We use standard quantisation for $\phi$, whose leading coefficient determines the BMN mass through \eqref{eq:massuplift}, and alternate quantisation for $v$, whose source is the appropriately rescaled finite part of $\Pi_v^{\rm ren}$. The Legendre term exchanges the boundary value of $v$ with that momentum:
\begin{equation}
 S_{\rm Leg}=-\frac{1}{2\kappa_2^2}
 \int_{\partial{\cal M}}\!{\rm d}t\,v\Pi_v^{\rm ren}.
 \label{eq:Legendre}
\end{equation}
Thus the bosonic action to be varied for arbitrary boundary time dependence is
\begin{equation}
 S_{\rm ren}=S_{\rm bulk}+S_{\rm GH}
 +S_{\rm ct}^{\rm susy}+S_{\rm ct}^{(2)}+S_{\rm Leg}.
 \label{eq:Sren}
\end{equation}
Replacing $W$ by $U_c$ changes the finite source definition. The BMN condition below selects the $W$ branch.

\subsection{Stationary BMN specialisation}\label{sec:stationary-renormalisation}

We now restrict to the radial, time-independent solutions of section~\ref{sec:boundarydata}. On the BPS branch the characteristic equations of $W$ are
\begin{align}
 \rho' &= 2e^Z W,
 &
 A' &= 2e^Z\partial_\rho W,
 \nonumber\\
 v' &= -\frac{4e^Z}{9\rho}\partial_v W,
 &
 \phi' &= -e^Z e^{-6v}\rho^{-1/3}\partial_\phi W.
 \label{eq:HJflow}
\end{align}
The same first-order system follows independently from the vanishing fermion variations of the gauged supergravity of \cite{Ortiz:2012ib}, restricted to our scalar sector and with the fermion signs specified in Appendix~\ref{app:killingspinors}. That appendix gives the projector, the explicit Killing spinor and the integrability check. The dilaton equation fixes $W=\frac12e^{-Z}\rho'$, identifying the Hamilton--Jacobi branch with the supersymmetric superpotential on this family.

\subsubsection{The BMN source condition}\label{sec:boundaryconditions}

For stationary solutions we use $w=r_+/r$ and the asymptotic expansion \eqref{eq:asymptoticqi}. The BMN deformation fixes the leading $\phi$ coefficient, while the independent ${\bf44}$ source is set to zero in the alternate quantisation of $v$. In the normalization used for the stationary solutions, this condition is
\begin{equation}
 {\cal J}_v\equiv\lim_{w\to0}\frac{w^2\Pi_v^{\rm ren}}{g^8r_+^7}=0.
 \label{eq:vsourcezero}
\end{equation}
The scalar fields behave as
\begin{align}
 v &=  a^3_0 w^2  -\frac{1}{2}\left(a^3_0\right)^2w^4  +a^3_3w^5+\ldots ,
 \label{eq:vasymptotic}
 \\
 \frac{\phi}{\rho^{1/3}} &=  \bar\mu\,w^{5/2}  +a^4_2w^{9/2}  +\ldots .
 \label{eq:phiasymptotic}
\end{align}
Substituting these expressions into \eqref{eq:Pivren}, the divergences in $w^2\Pi_v^{\rm ren}$ cancel between the canonical momentum and the counterterm contribution. The finite source associated to $v$ is, up to an overall nonzero normalisation,
\begin{equation}
 {\cal J}_v
 \propto
 \frac{27}{4}a^3_3-\bar\mu^2 .
 \label{eq:vsource}
\end{equation}
More generally, using $U_c$ in \eqref{eq:countertermc} changes the finite mixed contribution and gives
\begin{equation}
 {\cal J}_v^{(c)}\propto\frac{27}{4}a^3_3-c\bar\mu^2.
 \label{eq:vsourcec}
\end{equation}
The proportionality factor is the same as in \eqref{eq:vsource}. Independently, the matter-fermion equation fixes $a^3_3=4\bar\mu^2/27$ on the BPS family, as shown in Appendix~\ref{app:BPSsource}. Requiring the independent ${\bf44}$ source to vanish on that family for nonzero mass therefore fixes $c=1$ within \eqref{eq:countertermclass}. With this choice, the BMN source condition is
\begin{equation}
 a^3_3=\frac{4}{27}\bar\mu^2 .
 \label{eq:susybc}
\end{equation}
Equation~\eqref{eq:susybc} is not a condition on the coefficient $a^3_3$ alone: it states that the renormalised source in the alternate quantisation vanishes. The $\bar\mu^2$ term originates from the finite mixed counterterm in \eqref{eq:finitecounterterm}.

Equation \eqref{eq:BMNaction} makes clear why the ${\bf 44}$ and ${\bf 84}$ sectors cannot be treated as independent deformations. The terms associated with the ${\bf 84}$ are linear in $\mu$, whereas the bosonic mass deformation contains a ${\bf 44}$ component proportional to $\mu^2$. Supersymmetry ties these terms together into a single deformation. On the gravity side this statement is encoded in the renormalised source
\begin{equation}
 {\cal J}_{44}  \propto  a^3_3-\frac{4}{27}\bar\mu^2 .
 \label{eq:J44}
\end{equation}
Thus, setting ${\cal J}_{44}=0$ does not mean that the coefficient $a^3_3$ itself vanishes. Rather, it states that there is no \emph{independent} source for the ${\bf 44}$ operator beyond the one required by the supersymmetric completion of the BMN deformation. This is precisely the content of \eqref{eq:susybc}.

The determination of $c$ could be strengthened by including the fermions in the boundary variational problem. This would require the fermionic boundary terms and the source multiplet associated with the alternate quantisation of $v$, followed by a classification of finite supersymmetric boundary terms. We leave this for future work; the argument above fixes $c$ within \eqref{eq:countertermclass}, which is sufficient for the observables considered here.

\subsubsection{One-point functions}\label{sec:holoenergy}

For time-independent sources, $S_{\rm ct}^{(2)}=0$. The one-point functions follow by varying the renormalised action with
respect to the corresponding sources. In particular, varying the
boundary metric defines the energy, whereas varying the BMN mass
parameter defines its conjugate quantity $\Pi$. We write the on-shell
variation in the form
\begin{equation}
 \delta S_{\rm ren}\big|_{\rm EOM}
 =
 \int_{\partial{\cal M}}{\rm d}t\,
 \left[
 \frac{1}{2}\sqrt{-\gamma}\,
 {\cal T}^{tt}\delta\gamma_{tt}
 -\Pi\,\delta\mu
 -{\cal V}_{44}\,\delta{\cal J}_{44}
 \right],
 \label{eq:renvariation}
\end{equation}
where ${\cal V}_{44}$ denotes the expectation value conjugate to the
alternate-quantisation source. The mass-source term carries a minus sign because this is the Lorentzian action, whereas $\Pi$ is defined by the mass derivative of the Hamiltonian. In the ensemble relevant for the BMN
model we impose ${\cal J}_{44}=0$ and vary only $\mu$.

Evaluating \eqref{eq:renvariation} on the asymptotic expansion
\eqref{eq:asymptoticqi}, and imposing
\eqref{eq:susybc}, gives the following results, which we quote in units $2\kappa_2^2=1$; in these units the entropy \eqref{eq:entropy} below reads $S=\pi\rho(r_+)$. In the scalar variation we use $\delta m_\phi=\delta\mu/(4g^2)$, as required by \eqref{eq:massuplift}:
\begin{subequations}
\label{eqs:holothermo}
\begin{align}
 E &= \frac{g^8r_+^7}{840} \left( 945 -3591a^1_7 -2248\bar\mu^2a^3_0 -888\bar\mu a^4_2 \right),
 \label{eq:holoenergy}
 \\
 \Pi
 &= -\frac{1}{2}g^{9/2}r_+^{9/2} \left( 2\bar\mu a^3_0+a^4_2 \right).
 \label{eq:holopi}
\end{align}
\end{subequations}
These expressions agree precisely with those obtained from the
covariant phase space construction described in
Appendix~\ref{app:wald}.

There is an important conceptual difference between the two derivations. The covariant phase space argument determines \eqref{eqs:holothermo} only up to the variation of a local function of the source $\mu$. In holographic renormalisation this ambiguity is the usual freedom to add finite local counterterms. Within the class \eqref{eq:countertermclass}, the BMN source condition on the BPS family fixes the finite mixed term, and the prescription generated by $W$ gives \eqref{eqs:holothermo}. If a wider class of source-only finite terms is admitted, the energy can still be shifted by a function of $\mu$, with the corresponding shift in $\Pi$. We fix this remaining normalisation by requiring $E_{\rm BPS}(\mu)=0$ throughout the BPS family. This is a normalisation of the supersymmetric Hamiltonian; it does not follow from the bosonic Hamilton--Jacobi equation alone.

The quantity $\Pi$ also has a simple interpretation in the matrix quantum mechanics. Since $\mu$ is the single parameter entering the BMN deformation, the gravitational one-point function \eqref{eq:holopi} computes the thermal expectation value of the operator conjugate to $\mu$. Equivalently,
\begin{equation}
 \Pi = \frac{\partial F}{\partial\mu} = \left\langle \frac{\partial H_{\rm BMN}}{\partial\mu} \right\rangle ,
 \label{eq:PiMatrix}
\end{equation}
with the second equality understood in the thermal matrix model. Thus $\Pi$ combines the expectation values of the bosonic mass operators, the fermion bilinear and the Myers operator appearing in \eqref{eq:BMNaction}. This is the matrix-model counterpart of the fact that the asymptotic supergravity data $a^3_0$ and $a^4_2$ enter \eqref{eq:holopi} only through the supersymmetric combination $2\bar\mu a^3_0+a^4_2$.

\subsubsection{Thermodynamic relations and normalised observables}\label{sec:thermodynamics}

Under $t=-i\tau$, we define $I_{\rm ren}^{(E)}=-iS_{\rm ren}|_{t=-i\tau}$. The induced Euclidean metric is positive and its measure is $\sqrt{\gamma_E}$. The Euclidean renormalised on-shell action determines the free energy,
\begin{equation}
 I_{\rm ren}^{(E)}=\beta F .
 \label{eq:EuclideanFreeEnergy}
\end{equation}
Its variation at fixed temperature gives
\begin{equation}
 {\rm d}F=\Pi\,{\rm d}\mu ,
 \label{eq:firstlawFdimensionful}
\end{equation}
while the energy and entropy satisfy
\begin{equation}
 {\rm d}E=T\,{\rm d}S+\Pi\,{\rm d}\mu .
 \label{eq:firstlawdimensionful}
\end{equation}
Notice that this convention is equivalent to
${\rm d}E-\Pi\,{\rm d}\mu=T\,{\rm d}S$, which is the form obtained
directly from the covariant phase space analysis.

For a regular horizon at $w=1$ in \eqref{eq:staticparametrisation}, the Hawking temperature follows from Euclidean regularity and is
\begin{equation}
 T=\frac{7g^{7/2}r_+^{5/2}}{4\pi}.
 \label{eq:temperature}
\end{equation}
Together with \eqref{eq:mudef}, this gives $\widehat\mu=16\pi\bar\mu/7$ and $\widehat\mu_{\rm CGPS}=4\bar\mu/3$.
The entropy is determined by the Wald formula. Since the coefficient
of the Ricci scalar in \eqref{eq:actionbmn} is proportional to the
dilaton, one finds
\begin{equation}
 S
 =
 \frac{\pi}{2\kappa_2^2}\rho(r_+).
 \label{eq:entropy}
\end{equation}
If we identify $\kappa_2^2=2\pi G_2$, this becomes simply
\begin{equation}
 S=\frac{\rho(r_+)}{4G_2}.
 \label{eq:entropyG2}
\end{equation}

Finally, the reduced equations are invariant under a rescaling of the
horizon scale,
\begin{equation}
 r_+\longrightarrow\lambda r_+ .
\end{equation}
Under this transformation
\begin{equation}
 S\longrightarrow\lambda^{9/2}S, \qquad T\longrightarrow\lambda^{5/2}T, \qquad \mu\longrightarrow\lambda^{5/2}\mu, \qquad E\longrightarrow\lambda^7 E, \qquad \Pi\longrightarrow\lambda^{9/2}\Pi .
 \label{eq:thermoscaling}
\end{equation}
Euler's theorem, together with \eqref{eq:firstlawdimensionful}, then
gives the Smarr relation
\begin{equation}
 E=\frac{9}{14}TS+\frac{5}{14}\Pi\mu .
 \label{eq:Smarr}
\end{equation}
It is therefore natural to introduce
\begin{equation}
 \widehat E=\frac{E}{T^{14/5}}, \qquad \widehat S=\frac{S}{T^{9/5}}, \qquad \widehat\Pi=\frac{\Pi}{T^{9/5}}, \qquad \widehat\mu=\frac{\mu}{T}.
 \label{eq:dimensionlessquantities}
\end{equation}
In terms of these quantities,
\begin{subequations}
\begin{align}
 \widehat E
 &= \frac{9}{14}\widehat S +\frac{5}{14}\widehat\Pi\widehat\mu,
 \\
 {\rm d}\widehat E &= {\rm d}\widehat S +\widehat\Pi\,{\rm d}\widehat\mu .
\end{align}
\end{subequations}
For
\begin{equation}
 \widehat F=\widehat E-\widehat S
\end{equation}
we consequently obtain
\begin{subequations}
\begin{align}
 {\rm d}\widehat F &= \widehat\Pi\,{\rm d}\widehat\mu,
 \label{eq:dFhat}
 \\
 \widehat F&= \frac{5}{14} \left( \widehat\Pi\widehat\mu-\widehat S \right).
 \label{eq:FhatSmarr}
\end{align}
\end{subequations}

To make contact with previous work \cite{Costa:2014wya}, define further
\begin{equation}
\tilde{F}=\frac{\widehat{F}}{\widehat{F}|_{\mu=0}}\,,\quad\text{and}\quad\tilde{S}=\frac{\widehat{S}}{\widehat{S}|_{\mu=0}}\,,
\end{equation}
where both $\widehat{F}|_{\mu=0}$ and $\widehat{S}|_{\mu=0}$ can be computed from the D0 brane solution
\begin{equation}
\widehat{F}|_{\mu=0}=-\frac{5}{14}\widehat{S}|_{\mu=0}\,.
\end{equation}
Using the Smarr relation and the first law of black hole mechanics gives an additional numerical check
\begin{equation}
\left(1-\frac{5}{14}\widehat{\mu}\frac{\mathrm{d}}{\mathrm{d}\widehat{\mu}}\right)\tilde{F}=\tilde{S}\,,
\label{eq:firstlaw}
\end{equation}
which was first derived in \cite{Costa:2014wya}, where it was used to determine $\tilde{F}$ from $\tilde{S}$. Here the energy and the mass response are extracted independently, and two distinct checks are available. Using the directly computed $\tilde\Pi$ in place of $-{\rm d}\tilde F/{\rm d}\widehat\mu$, \eqref{eq:firstlaw} becomes the algebraic Smarr relation \eqref{eq:FhatSmarr}, $\tilde S=\tilde F+\frac{5}{14}\widehat\mu\tilde\Pi$. This relation tests the boundary charges against the horizon entropy, but not the differential first law. The latter, ${\rm d}\tilde F/{\rm d}\widehat\mu=-\tilde\Pi$, can be tested separately by integrating $\tilde\Pi$ along the branch.

\section{Solutions of the two-dimensional model}\label{sec:solutions}

We now turn to solutions of the reduced equations. The undeformed D0 solution provides the starting point for the thermal branch. We also describe an analytic supersymmetric solution and check its charges in the scheme of section~\ref{sec:renormalisation}, before formulating the thermal boundary-value problem.

\subsection{The undeformed D0 solution}\label{sec:D0solution}

At $\bar\mu=0$, the non-extremal D0 solution is recovered by setting
\begin{equation}
 q_1=q_2=1,\qquad q_3=q_4=0.
 \label{eq:D0qfunctions}
\end{equation}
Thus, $v=\phi=0$, $\rho=(gr)^{9/2}$, and
\begin{equation}
 {\rm d}s_2^2=-(gr)^7\tilde{g}(r){\rm d}t^2 +\frac{{\rm d}r^2}{\tilde{g}(r)}\quad\text{with}\quad \tilde{g}(r)=1-\left(\frac{r_+}{r}\right)^7\,.
 \label{eq:D0twodimensional}
\end{equation}
Its uplift gives the undeformed geometry described in section~\ref{sec:gravitydual}, after a choice of coordinates and circle normalisation. This solution fixes the reference values used in the normalised thermodynamic quantities of section~\ref{sec:thermodynamics}.

\subsection{An exact supersymmetric solution}\label{sec:susy}

\subsubsection{First-order flow and asymptotic checks}

We obtain the first-order system by setting the fermionic supersymmetry variations of \cite{Ortiz:2012ib} to zero in our truncation. The fermion signs, their relation to the printed formulas, and the conversion to our Clifford-algebra conventions are given explicitly in Appendix~\ref{app:killingspinors}. The Killing spinor conditions are
\begin{equation}
 \delta_\epsilon\psi_\mu^I=0,\qquad
 \delta_\epsilon\psi_2^I=0,\qquad
 \delta_\epsilon\chi^{aI}=0,
 \label{eq:KillingSpinorConditions}
\end{equation}
where $\psi_\mu^I$, $\psi_2^I$ and $\chi^{aI}$ denote the gravitini, dilatini and matter fermions in the notation of that reference. For the static Ansatz, the resulting scalar and metric equations on the branch approaching the D0 boundary are
\begin{subequations}
\label{eq:firstordersystem}
\begin{align}
&\rho^\prime = \frac{3}{2}\,g\,\Xi\,\rho ^{7/9} e^{Z-2 v}\,,
\label{eq:firstorderrho}
\\
&\phi^\prime = -\frac{3\,g\,\rho^{-2/9}}{\Xi }\,e^{Z-2 v}\phi  \,,
\\
&v^\prime = \frac{2}{3}\frac{g}{\Xi  \,\rho ^{8/9}} \left[\left(1+e^{3 v}-2 e^{6 v}\right) \rho ^{2/3}-2 e^{6 v} \phi ^2\right]e^{Z-2 v} \,,
\\
&A^\prime = \frac{1}{6}\frac{g}{\Xi \,\rho ^{8/9}}\left[7 \left(1+2 e^{3 v}\right)^2 \rho ^{2/3}+16 e^{6 v} \phi ^2\right]e^{Z-2 v}\,,
\label{eq:firstorder}
\end{align}
where
\begin{equation}
\Xi\equiv \sqrt{(1+2 e^{3 v})^2+\frac{4 e^{6 v} \phi ^2}{\rho ^{2/3} }}\,.
\end{equation}
\end{subequations}

These equations agree with the Hamilton--Jacobi flow \eqref{eq:HJflow}. Their derivation from \eqref{eq:KillingSpinorConditions} supplies the independent supersymmetry input used to identify $W$ and fix the finite mixed counterterm in section~\ref{sec:boundaryconditions}.

Expressing this flow in the asymptotic parametrisation \eqref{eq:staticparametrisation} and expanding near the boundary gives \eqref{eq:susybc}. The finite part of the momentum conjugate to $\phi$ is proportional to
\begin{equation}
 \Pi_\phi^{\rm ren}  \propto  a^4_2+2\bar\mu a^3_0 .
 \label{eq:phimomentumfinite}
\end{equation}
For a supersymmetric solution this vanishes, and hence
\begin{equation}
 a^4_2=-2\bar\mu a^3_0 ,
 \label{eq:susyphivev}
\end{equation}
in agreement with the direct expansion of the first-order equations. The remaining supersymmetric relation is
\begin{equation}
 a^1_7 = \frac{5}{19} -\frac{472}{3591}\bar\mu^2a^3_0 .
 \label{eq:susymetricvev}
\end{equation}

Substituting \eqref{eq:susyphivev} and \eqref{eq:susymetricvev} into \eqref{eqs:holothermo} gives
\begin{equation}
 E_{\rm BPS}=0,
 \qquad
 \Pi_{\rm BPS}=0.
 \label{eq:BPScharges}
\end{equation}
The counterterm prescription generated by $W$ therefore implements
both the BMN source condition \eqref{eq:susybc} and the chosen
normalisation of the supersymmetric Hamiltonian.

\subsubsection{Integration of the flow}

Let us fix the gauge by taking
\begin{equation}
Z=2 v+\frac{2}{9}\log \rho +\log \Xi\,,
\end{equation}
and
\begin{equation}
\phi=\rho^{1/3}\sqrt{\eta}.
\end{equation}
so that Eqs.~\eqref{eq:firstordersystem} can be written as
\begin{subequations}
\begin{align}
&\frac{\rho^\prime}{\rho}=\frac{3}{2}\,g\,\Xi^2\,,
\\
& \frac{\eta^\prime}{\eta}=-g(6+\Xi^2)\,,\label{eq:middlem}
\\
&v^\prime=\frac{2}{3}g(1+e^{3v}-2 e^{6v}-2 e^{6v}\eta)\,,\label{eq:middle}
\\
& A^\prime=g\,\left(\frac{7}{6}\Xi^2-2e^{6v}\eta\right)\,,
\end{align}
where now
\begin{equation}
\Xi=\sqrt{(1+2 e^{3 v})^2+4 e^{6 v} \eta}
\end{equation}
\end{subequations}

One can use Eq.~(\ref{eq:middle}) to rewrite Eq.~(\ref{eq:middlem}) as a second order equation for $v$. This procedure turns out to yield
\begin{equation}
v^{\prime\prime}-9 {v^{\prime}}^2+(15+10 e^{3v}-4 e^{6v})g v^\prime-2(3+5e^{3v}-4 e^{6v}-4 e^{9v})g^2=0\,.
\end{equation}

Finally, we introduce the dimensionless radial coordinate $z\equiv r g$, so that the dependence in $g$ of the above equation cancels out and we are left with
\begin{equation}
\ddot{v}-9 \dot{v}^2+(15+10 e^{3v}-4 e^{6v}) \dot{v}-2(3+5e^{3v}-4 e^{6v}-4 e^{9v})=0\,,
\end{equation}
where $\dot{}$ denotes derivatives with respect to $z$. It is still possible to reduce the above to a first-order equation. We first note that in the above equation there is no explicit $z$ dependence. As such we can regard $\dot{v}$ as a function of $v$ only. In fact, since the only dependence in $v$ appears through multiples of $e^{3v}$, we set $\dot{v}=P(e^{3v})$, which brings the above equation to the following form
\begin{equation}
\frac{\mathrm{d}P}{\mathrm{d}y}=\frac{6+10 y-8 y^2-8y^3-(15+10y-4y^2)P+9P^2}{3 y P}\,.
\label{eq:Pequation}
\end{equation}
where $y=e^{3v}$. Given $P$, Eq.~(\ref{eq:middle}) determines the remaining scalar,
\begin{equation}
\eta=\frac{1+y-2y^2-\frac{3}{2}P}{2y^2}\,,
\label{eq:etaofP}
\end{equation}
so that each solution of \eqref{eq:Pequation} is a trajectory in the $(y,\eta)$ plane.

We have found a few solutions to the above differential equation, but we have not managed to find the generic solution (which should depend on one integration constant). Only one of the solutions we found generates a non-vanishing profile for $\rho$, $\eta$ and $v$, which corresponds to
\begin{equation}
P=1+\frac{5}{3}y-\frac{4}{3}y^2-\frac{4}{3}y^3\,.
\label{eq:Pexplicit}
\end{equation}
We integrate this solution explicitly in Appendix~\ref{app:flowintegration}. It approaches the D0 boundary with $\phi\propto R^{-1}$ and $v\propto R^{-5}$, as shown in \eqref{eq:BPSvasymptotic}, so that $a^3_0=0$. Its integration constant $z_0$ fixes the BMN mass. As $z\to z_0$ the flow becomes singular: in terms of the proper radial distance $\chi$ from $z=z_0$, $\rho\propto\chi^{9/7}$, $\phi$ tends to a constant and $v\simeq-\frac{2}{7}\log\chi$.

The region $z\to z_0$ is the large-$y$ region of the flow. There \eqref{eq:etaofP} gives $\eta\simeq y$, and this behaviour is not special to \eqref{eq:Pexplicit}. For $y\gg1$ the first-order equations reduce to ${\rm d}\eta/{\rm d}v\simeq3\eta$, so that every trajectory with $\eta\neq0$ behaves as
\begin{equation}
\eta\simeq C\,y\,,\qquad y\gg1\,,
\label{eq:largeyC}
\end{equation}
for some constant $C$, with $C=1$ for \eqref{eq:Pexplicit}. The opposite end of this family is the solution
\begin{equation}
P_\star=\frac{2}{3}\left(1+y-2y^2\right)\,,
\label{eq:Pstar}
\end{equation}
which solves \eqref{eq:Pequation} and has $\eta=0$ by \eqref{eq:etaofP}. It describes a supersymmetric flow of the ${\bf 44}$ scalar $v$ alone, with $\phi=0$. It therefore does not satisfy the BMN boundary conditions by itself. We will nevertheless find in section~\ref{sec:susylimit} that it controls the end of the thermal branch.

\subsection{Thermal boundary-value problem}\label{sec:thermal}

For finite temperature we use the parametrisation \eqref{eq:staticparametrisation} on $0\leq w\leq1$, with the boundary at $w=0$ and the horizon at $w=1$. Regular, nonzero metric functions at $w=1$ give a non-extremal horizon. Substituting this Ansatz into \eqref{eqs:truncation} gives three second-order equations for $q_2$, $q_3$ and $q_4$, and a first-order equation for $q_1$. The powers of $g$ and $r_+$ in the parametrisation remove both scales from the equations for the $q_i$.

At the boundary, we impose the D0 asymptotics and fix $q_4(0)=\bar\mu$. We also impose the BMN source condition \eqref{eq:susybc}. The latter specifies the deformation of the theory; it does not require the thermal solution to satisfy the supersymmetric first-order equations.

At the black hole horizon we assume regularity in ingoing coordinates across the null hypersurface $r=r_+$, which amounts to an expansion for the $q_i$ of the form
\begin{equation}
q_i(w) = \sum_{j=0}^{+\infty}b^i_j(1-w)^{j}\,.
\end{equation}
Solving the equations of motion order by order shows that the free data on the horizon can now be parametrised by four integration constants $b^1_0$, $b^2_0$, $b^3_0$ and $b^4_0$.

\subsubsection{Numerical variables and boundary conditions}

For the numerical calculation, it is convenient to factor out the leading asymptotic behaviour:
\begin{equation}
q_1=1+w^4\,r_1\,,\quad q_2=1+w^4\,r_2\,,\quad q_3=r_3\quad \text{and}\quad q_4=r_4\,.
\end{equation}
The boundary conditions at the conformal boundary now become simply
\begin{equation}
r_1(0)+\frac{14}{13}r_3(0)^2=0\,,\quad r_2(0)+\frac{2}{13}r_3(0)^2=0\,,\quad \left.\frac{\mathrm{d}r_3}{\mathrm{d}w}\right|_{w=0}=0\quad\text{and}\quad r_4(0)=\bar{\mu}.
\end{equation}

At the horizon, located at $w=1$, we have three Robin boundary conditions for $r_1$, $r_3$ and $r_4$ and a Dirichlet condition $r_2(1)=r_2^{\mathcal{H}}$, where $r_2^{\mathcal{H}}$ denotes the horizon value of the numerical function $r_2$. This value is determined as part of the solution. For each fixed $\bar\mu$, we adjust it to impose the BMN source condition \eqref{eq:susybc}, which becomes
\begin{equation}
\left.\frac{\mathrm{d}^3r_3}{\mathrm{d}w^3}\right|_{w=0}=\frac{8}{9}\bar{\mu}^2\,.
\label{eq:susy}
\end{equation}
Thus, for each value of $\bar{\mu}$ we have the required number of conditions to solve the boundary value problem. In terms of constants, the local solutions of the four equations form a seven-parameter family. Regularity at the horizon leaves the four constants $b^i_0$, while the D0 asymptotics leave the five constants of section~\ref{sec:boundarydata}. Solutions regular at both ends therefore form a two-parameter family, which we may label by $\bar\mu$ and $a^3_3$, and the BMN source condition \eqref{eq:susy} leaves a single solution for each $\bar\mu$. The horizon value $q_2^{\mathcal{H}}\equiv q_2(1)=1+r_2^{\mathcal{H}}$ fixes the entropy: by \eqref{eq:entropy} and \eqref{eq:temperature}, $\tilde S=(q_2^{\mathcal{H}})^{9/2}$.

\subsubsection{Continuation to the end of the branch}\label{sec:continuation}

We solve the boundary-value problem by Newton's method on Chebyshev--Lobatto grids, starting from the D0 solution and increasing $\bar\mu$ in small steps. This works well until $\widehat\mu$ approaches $4\pi$. There $q_2^{\mathcal{H}}$ decreases rapidly while $\bar\mu$ barely changes, ${\rm d}\bar\mu/{\rm d}q_2^{\mathcal{H}}\to0$, and a sequence of fixed increments in $\bar\mu$ can no longer reach the end of the branch. We therefore exchange the roles of the two parameters. We prescribe $q_2^{\mathcal{H}}$ and treat $\bar\mu$ as an unknown, with $r_4(0)=\bar\mu$ and \eqref{eq:susy} providing the two conditions that involve it. Since the bulk equations do not contain $\bar\mu$ explicitly, the corresponding column of the Jacobian is nonzero only in the rows associated with these two conditions. We decrease $q_2^{\mathcal{H}}$ geometrically and predict each new solution by polynomial extrapolation through the previous four.

The collocation Jacobian is severely ill-conditioned near the end of the branch; at $N=200$ its condition number is of order $10^{25}$. Factorisations in double precision followed by iterative refinement do not converge. Instead, we solve the linear systems by dense LU factorisation in 40-digit arithmetic, on grids with $N+1$ points for $N=200$, $300$ and $400$. At each step we monitor the Chebyshev coefficients of the $r_i$. As independent checks we use the Smarr relation \eqref{eq:FhatSmarr}, the integrated first law and the comparison between resolutions.

\section{Results}\label{sec:results}
\begin{figure}[t]
\centering
\includegraphics[width=\textwidth]{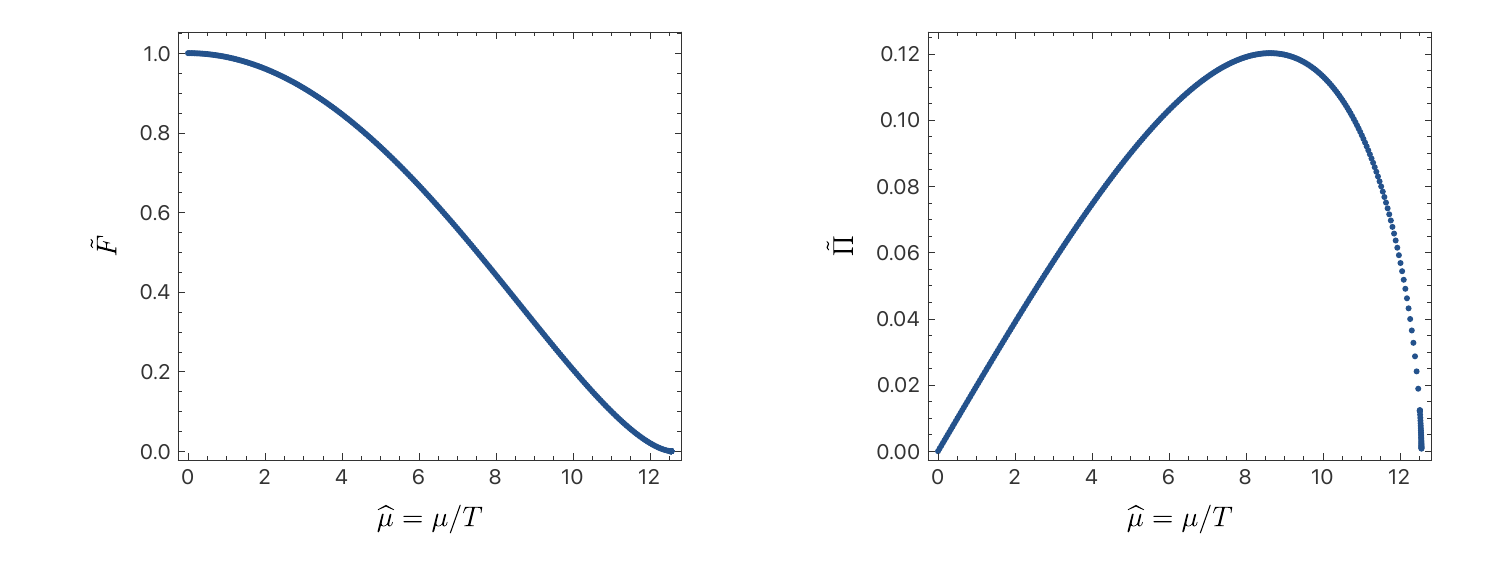}
\caption{Thermodynamics of the two-dimensional branch. Every point is a numerical solution; the points are not joined. The mass response is computed directly from \eqref{eq:holopi}, independently of the free-energy data.}
\label{fig:thermocurves}
\end{figure}

We display the thermodynamics in terms of the BMN mass parameter $\widehat\mu=\mu/T$. In addition to $\tilde F$ in \eqref{eq:firstlaw}, we define a dimensionless mass response by
\begin{equation}
 \tilde\Pi\equiv\frac{\widehat\Pi}{-\widehat F|_{\mu=0}}
 =-\frac{\mathrm d\tilde F}{\mathrm d\widehat\mu}.
 \label{eq:tildePi}
\end{equation}
The second equality follows from \eqref{eq:dFhat}; we do not use it to obtain the data in Fig.~\ref{fig:thermocurves}. Instead, we evaluate $\Pi$ directly from the boundary coefficients in \eqref{eq:holopi} at each of the 329 two-dimensional solutions. The normalised free energy $\tilde F$ decreases monotonically along the branch, so the physical free energy rises from a negative value towards zero. The response first grows and then tends back towards zero.

We can check the Smarr relation without differentiating the free-energy data. With the directly computed $\tilde\Pi$ it reads $\tilde S=\tilde F+(5/14)\widehat\mu\tilde\Pi$. Comparing the right-hand side, obtained from the asymptotic coefficients, with the entropy obtained at the horizon, we find that $100\bigl|\tilde F+(5/14)\widehat\mu\tilde\Pi-\tilde S\bigr|/\tilde S$ remains below $10^{-5}\%$ over all 329 solutions. As explained in section~\ref{sec:thermodynamics}, this tests the boundary charges against the horizon entropy rather than the differential first law.

The last point in Fig.~\ref{fig:thermocurves} lies at $\widehat\mu=12.56596$, close to $4\pi=12.56637$, with $\tilde F\simeq1.9\times10^{-7}$ and $\tilde\Pi\simeq7.6\times10^{-4}$. The conformal-coordinate data shown in Appendix~\ref{app:conformal} end at $\widehat\mu\simeq10.525$, where $\tilde F\simeq0.149$ and $\tilde S\simeq0.546$. An arclength continuation extends the conformal branch to $\widehat\mu\simeq11.22$, where its best-resolved solution agrees with the two-dimensional free energy to $6\times10^{-5}$.

\begin{figure}[t]
\centering
\includegraphics[width=\textwidth]{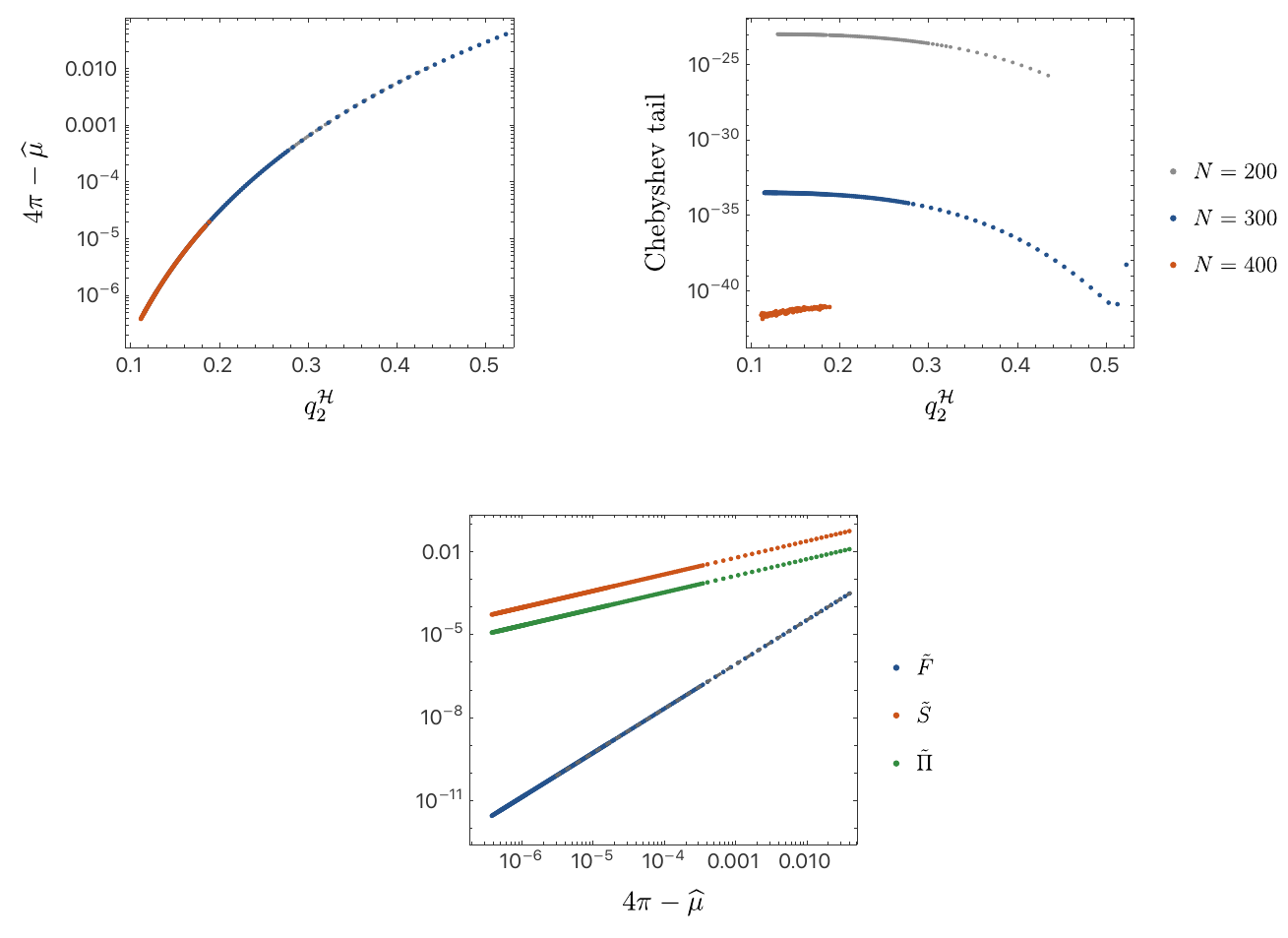}
\caption{The continuation in $q_2^{\mathcal{H}}$. Top left: distance of $\widehat\mu$ from $4\pi$ for three resolutions. Top right: largest relative size of the last tenth of the Chebyshev coefficients of the $r_i$, for the same resolutions. Bottom: $\tilde F$, $\tilde S$ and $\tilde\Pi$ against $4\pi-\widehat\mu$ for $N=300$ and $400$; the dashed line is a power-law fit to $\tilde F$ with exponent $1.6$. Every point is a numerical solution.}
\label{fig:nearend}
\end{figure}

Beyond the last point of Fig.~\ref{fig:thermocurves} we have continued the branch in $q_2^{\mathcal{H}}$, as described in section~\ref{sec:continuation}; see Fig.~\ref{fig:nearend}. Down to the smallest $q_2^{\mathcal{H}}$ we have reached, $\widehat\mu$ increases monotonically and we find no turning point. At $q_2^{\mathcal{H}}=0.1606$ and $N=400$, which we use as a reference point below and in Table~\ref{tab:susylimit}, we find
\begin{equation}
 4\pi-\widehat\mu\simeq5.8\times10^{-6},\qquad
 \tilde F\simeq2.2\times10^{-10},\qquad
 \tilde S\simeq2.7\times10^{-4},\qquad
 \tilde\Pi\simeq5.9\times10^{-5},
 \label{eq:lastpoint}
\end{equation}
and the Smarr relation holds to a relative accuracy better than $10^{-18}$. To test the first law itself we compare $\tilde F(q_2^{\mathcal{H}})-\tilde F(q_{2,0}^{\mathcal{H}})$ with $-\int\tilde\Pi\,{\rm d}\widehat\mu$, integrating a high-order interpolant of the directly computed $\tilde\Pi$ along each continuation. The two agree to a relative accuracy of $10^{-10}$ for $N=300$, over $0.16\lesssim q_2^{\mathcal{H}}\leq0.52$, and of $10^{-9}$ for $N=400$; in both cases the residual is set by the spacing of the solutions rather than by their accuracy. The solutions with $N=200$ and $N=400$ agree in $\bar\mu$ to ten significant figures at $q_2^{\mathcal{H}}=0.1848$ and $0.1714$. At $N=200$, however, the Smarr relation is satisfied only to $10^{-8}$ by $q_2^{\mathcal{H}}\simeq0.18$, and the continuation becomes unreliable below $q_2^{\mathcal{H}}\simeq0.12$; we quote only data with $N=300$ and $400$ there. At $N=300$ the Smarr relation holds to $10^{-18}$ down to $q_2^{\mathcal{H}}\simeq0.12$, but it degrades rapidly below $q_2^{\mathcal{H}}\simeq0.115$, and we do not use those solutions. Where both resolutions are available, $0.112\lesssim q_2^{\mathcal{H}}\leq0.189$, the $N=300$ and $N=400$ solutions agree in $\bar\mu$, $\tilde S$ and $\tilde\Pi$ to $10^{-16}$.

Near the end of the branch the thermodynamic quantities follow power laws accurately. We fit $\log\tilde F$, $\log\tilde S$ and $\log\tilde\Pi$ against $\log(\widehat\mu_c-\widehat\mu)$ in windows of $4\pi-\widehat\mu$, leaving $\widehat\mu_c$ and the exponents free. Table~\ref{tab:fitstability} shows how the results depend on the window and the resolution. As the window moves towards the end point, the fitted $\widehat\mu_c$ approaches $4\pi$ monotonically and the exponents approach $1.6$ and $0.6$. The residual offsets decrease with the window in the way expected from corrections to scaling, and agree between $N=300$ and $N=400$. In the windows closest to the end point, $\widehat\mu_c-4\pi$ is $7\times10^{-13}$ from $\tilde F$ and $3\times10^{-12}$ from $\tilde S$ and $\tilde\Pi$. We therefore estimate $|\widehat\mu_c-4\pi|\lesssim10^{-10}$, which covers the spread between the three quantities and the drift between the two innermost windows. To the accuracy of our data,
\begin{equation}
 \tilde F\propto(\widehat\mu_c-\widehat\mu)^{8/5},\qquad
 \tilde S\propto\tilde\Pi\propto(\widehat\mu_c-\widehat\mu)^{3/5},\qquad
 \widehat\mu_c=4\pi ,
 \label{eq:endexponents}
\end{equation}

\begin{table}[t]
\centering
\small
\setlength{\tabcolsep}{4pt}
\begin{tabular}{llccccc}
\hline
 & window in $4\pi-\widehat\mu$ & points & $\widehat\mu_c-4\pi$ ($\tilde F$) & $\widehat\mu_c-4\pi$ ($\tilde S$) & $p_F$ & $p_S$\\
\hline
$N=300$ & $(10^{-4},10^{-2})$ & 33 & $1.4\times10^{-7}$ & $6.0\times10^{-7}$ & 1.60102 & 0.60165\\
 & $(10^{-5},10^{-3})$ & 59 & $2.0\times10^{-9}$ & $8.5\times10^{-9}$ & 1.600145 & 0.600234\\
 & $(3\times10^{-6},10^{-4})$ & 70 & $1.1\times10^{-10}$ & $4.8\times10^{-10}$ & 1.600034 & 0.600054\\
 & $(3\times10^{-6},1.2\times10^{-5})$ & 37 & $2.0\times10^{-11}$ & $8.7\times10^{-11}$ & 1.600012 & 0.600020\\
 & $(5\times10^{-7},3\times10^{-6})$ & 71 & $9.0\times10^{-13}$ & $3.9\times10^{-12}$ & 1.600003 & 0.600005\\
$N=400$ & $(3\times10^{-6},10^{-4})$ & 46 & $3.0\times10^{-11}$ & $1.3\times10^{-10}$ & 1.600015 & 0.600024\\
 & $(3\times10^{-6},1.2\times10^{-5})$ & 37 & $2.1\times10^{-11}$ & $8.9\times10^{-11}$ & 1.600012 & 0.600020\\
 & $(4\times10^{-7},3\times10^{-6})$ & 82 & $7.3\times10^{-13}$ & $3.2\times10^{-12}$ & 1.600003 & 0.600004\\
\hline
\end{tabular}
\caption{Stability of the power-law fits $\log X=c_X+p_X\log(\widehat\mu_c-\widehat\mu)$ near the end of the branch, with $\widehat\mu_c$ and $p_X$ free. Fits to $\tilde\Pi$ give values of $\widehat\mu_c-4\pi$ between $3\%$ and $11\%$ below those from $\tilde S$, and exponents $p_\Pi$ below $p_S$ by at most $2\times10^{-4}$. For $N=300$ we use only solutions with $q_2^{\mathcal{H}}\geq0.115$; all $N=400$ solutions lie in $3.8\times10^{-7}<4\pi-\widehat\mu<1.9\times10^{-5}$.}
\label{tab:fitstability}
\end{table}

These exponents are consistent with each other. The first law requires those of $\tilde F$ and $\tilde\Pi$ to differ by one. Since $\tilde F\ll\tilde S$ near the end point, the Smarr relation gives $\tilde S\simeq(5/14)\widehat\mu_c\tilde\Pi$; at the reference point $\tilde S/\tilde\Pi=4.488$, compared with $10\pi/7=4.488$.

The improved data also change the interpretation of the transition inferred in Ref.~\cite{Costa:2014wya}. At the value $T/\mu\simeq0.105905$ quoted there, both our two-dimensional and conformal-coordinate curves give $\tilde F\simeq0.270$, so the black-hole free energy remains negative. As discussed in Appendix~\ref{app:conformal}, the old crossing can be traced to the way the free energy was reconstructed from the entropy in Ref.~\cite{Costa:2014wya}. Integrating a polynomial fit of $\tilde S$ term by term strongly amplifies odd powers of $\widehat\mu$, and applying the same procedure to our entropy data produces spurious zeros around the quoted value, whereas restricting the fit to even powers reproduces the end point $T/\mu=1/(4\pi)$. Separately, the 2014 curve agrees with ours at small $\widehat\mu$ only after the empirical rescaling \eqref{eq:CGPSrescaling} of its horizontal axis, whose origin we have not identified. Provided this branch remains the dominant one up to its end point, there is nevertheless a transition, and it occurs at the end of the branch. With the normalisation $E_{\rm BPS}=0$ of section~\ref{sec:holoenergy}, the supersymmetric vacua \eqref{eq:BMNvacua} have zero energy, and the corresponding confined backgrounds have no horizon, so their free energy vanishes at order $N^2$. As $\widehat\mu\to\widehat\mu_c$, the free energy, entropy and mass response of the black holes tend to these values, with the exponents in \eqref{eq:endexponents}. The free energy and its first derivative are therefore continuous, while ${\rm d}^2\tilde F/{\rm d}\widehat\mu^2\propto(\widehat\mu_c-\widehat\mu)^{-2/5}$ diverges, so the transition appears to be continuous rather than first order. It coincides with the end of the branch, where the solutions become singular: as shown in section~\ref{sec:horizongeometry}, the curvature of the horizon at the pinch grows without bound as ${\cal R}_2(1)\to0$ at fixed temperature.

The same phenomenon was found for mass-deformed M2 branes in Stenzel space \cite{Dias:2017stenzel}. There, too, a family of black branes ends in a singular solution whose free energy equals that of the confined phase and approaches it with vanishing derivative, so that the first-order transition one might have expected coincides instead with the singularity. Since supergravity breaks down at the end point, the order of the transition and its exponents could be modified by stringy corrections there. The transition could also be pre-empted by a first-order transition to another branch of black holes with lower free energy, which our analysis cannot exclude. Lattice Monte Carlo calculations compare with $T/\mu\simeq0.106$ \cite{Schaich:2020ubh,Jha:2024rxz,Bergner:2021goh}. They approach this value as $g=\lambda/\mu^3$ grows towards $0.01$ \cite{Jha:2024rxz} and match it near $g\approx27$ at $N=12$ \cite{Bergner:2021goh}. At the smaller couplings, however, $T/\lambda^{1/3}\simeq0.5$, outside the supergravity regime $T,\mu\ll\lambda^{1/3}$; at the larger ones, finite-$N$ corrections are expected to be significant, and larger critical temperatures were indeed attributed to them. The agreement therefore does not directly test the large-$N$ supergravity value, which, if the branch studied here remains dominant to its end point, is $T/\mu=1/(4\pi)\simeq0.080$.

\subsection{Approach to a supersymmetric flow}\label{sec:susylimit}

Near the end of the branch the mass response and the entropy decrease towards zero, but this does not mean that the solutions approach the explicit supersymmetric flow of section~\ref{sec:susy}. That flow has $a^3_0=0$, since the large-$R$ expansion \eqref{eq:BPSvasymptotic} of $v$ starts at order $R^{-5}$, whereas $a^3_0$ multiplies the $w^2$ term in \eqref{eq:vasymptotic}. In the thermal data, by contrast, $a^3_0=q_3(0)$ grows from $3.14$ at $\widehat\mu=12.527$ to about $110$ at the reference point. The vanishing of $\Pi$ comes from a cancellation between $2\bar\mu a^3_0$ and $a^4_2$, rather than from either coefficient tending to zero.

\begin{figure}[t]
\centering
\includegraphics[width=\textwidth]{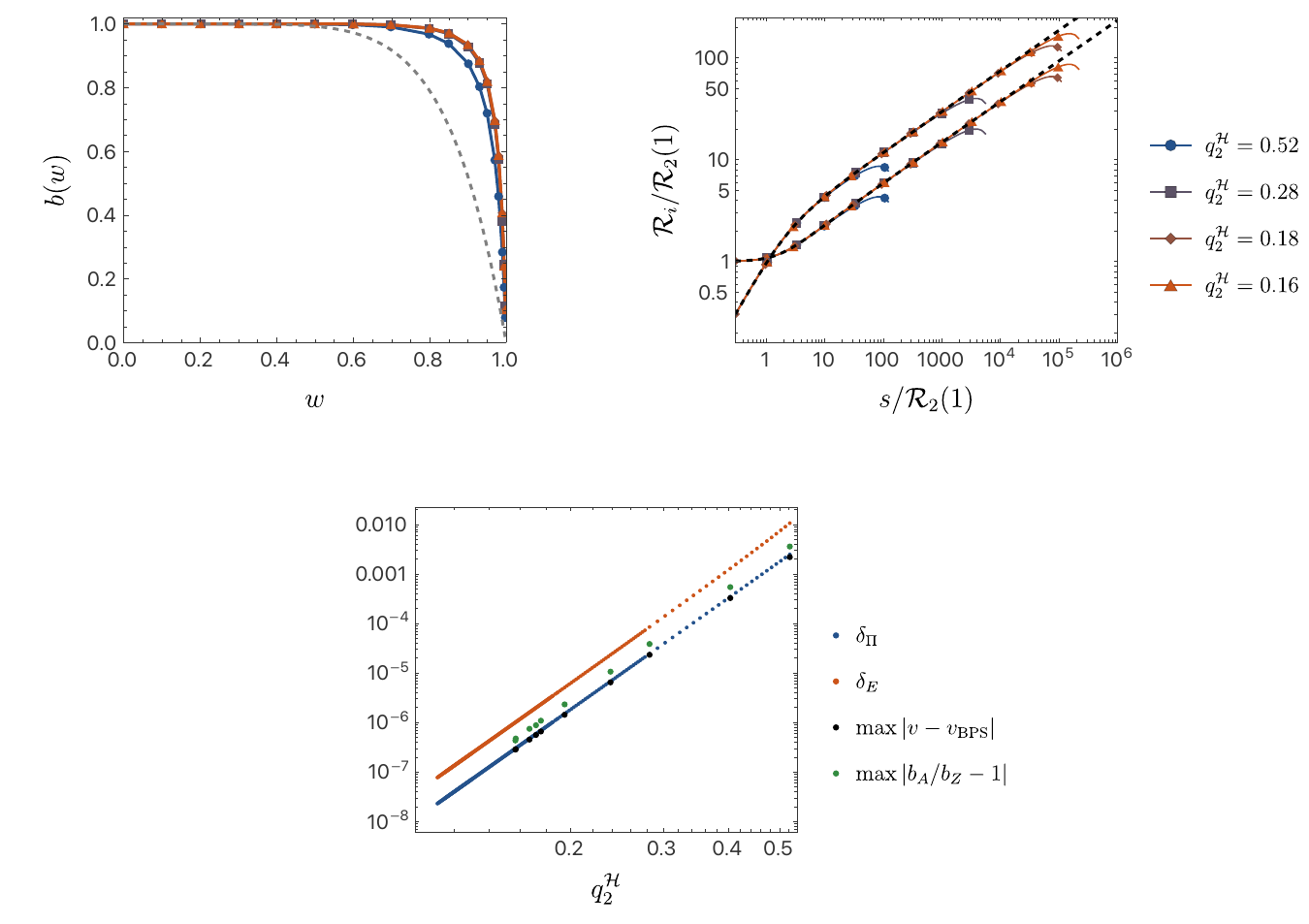}
\caption{Top left: the blackening function $b(w)$ at four points along the branch; the dashed curve is $1-w^7$. Top right: radii of the $S^2$ and $S^5$ along the horizon from $\zeta=1$, in units of the neck, against the proper distance $s$, for the same four solutions. The dashed curves are the analytic profile \eqref{eq:neckprofile} with $\eta_{\cal H}=1$; the numerical curves leave it where $\zeta$ departs from one. Bottom: deviation of the thermal solutions from the supersymmetric relations \eqref{eq:susyphivev} and \eqref{eq:susymetricvev}, from the BPS trajectory \eqref{eq:flowlogrho}, and between the two blackening factors in \eqref{eq:blackenedBPS}, as functions of $q_2^{\mathcal{H}}$.}
\label{fig:bps}
\end{figure}

This cancellation is precisely the supersymmetric relation \eqref{eq:susyphivev}. The second relation, \eqref{eq:susymetricvev}, is approached in the same way, as shown in Table~\ref{tab:susylimit}. Since \eqref{eqs:holothermo} vanishes when both relations hold, the asymptotic coefficients which enter $E$ and $\Pi$ approach those of a supersymmetric solution with $a^3_0\neq0$, for which \eqref{eq:BPScharges} holds. Their deviation from these relations is shown along the branch in Fig.~\ref{fig:bps}.

We can test the bulk solutions against the first-order equations directly. Dividing \eqref{eq:firstordersystem} by \eqref{eq:firstorderrho} eliminates $Z$. In terms of $y=e^{3v}$ and $\eta=\phi^2\rho^{-2/3}$ we find
\begin{subequations}
\label{eq:flowlogrho}
\begin{align}
 \frac{{\rm d}v}{{\rm d}\log\rho}&=\frac{4}{9}\,\frac{1+y-2y^2-2y^2\eta}{\Xi^2}\,,
 \\
 \frac{{\rm d}\eta}{{\rm d}\log\rho}&=-\eta\left(\frac{2}{3}+\frac{4}{\Xi^2}\right),
 \\
 \frac{{\rm d}A}{{\rm d}\log\rho}&=\frac{1}{9}\,\frac{7(1+2y)^2+16y^2\eta}{\Xi^2}\,,
\end{align}
\end{subequations}
with $\Xi^2=(1+2y)^2+4y^2\eta$. The first two equations do not involve the radial gauge. Their trajectories in the $(y,\eta)$ plane are the solutions of \eqref{eq:Pequation}, through \eqref{eq:etaofP}. We integrate them from the values of $v$ and $\eta$ of the thermal solution at $w=3/10$, in both directions. The thermal $v$ and $\eta$, regarded as functions of $\log\rho$, then lie on this BPS trajectory all the way from the boundary to the horizon, to the accuracy shown in Table~\ref{tab:susylimit}.

The metric, on the other hand, is not supersymmetric. In the radial coordinate of \eqref{eq:staticparametrisation}, we define $A_{\rm BPS}$ by integrating the last equation in \eqref{eq:flowlogrho} along the solution, normalised at the boundary, and $Z_{\rm BPS}$ from \eqref{eq:firstorderrho}. We then find
\begin{equation}
 {\rm d}s_2^2=-b(r)\,e^{2A_{\rm BPS}}{\rm d}t^2+\frac{e^{2Z_{\rm BPS}}}{b(r)}\,{\rm d}r^2\,.
 \label{eq:blackenedBPS}
\end{equation}
The two determinations of $b$, from $A$ and from $Z$, are independent. They agree to the accuracy shown in the last column of Table~\ref{tab:susylimit}. The blackening function vanishes linearly at the horizon and tends to one at the boundary. It is neither $1-w^7$ nor $1-(\rho_{\cal H}/\rho)^{14/9}$, and it has converged along the branch; for instance, $b(0.8)=0.98642$ at $q_2^{\mathcal{H}}=0.28$ and $0.98652$ at the reference point; see Fig.~\ref{fig:bps}. At fixed radial position $b$ therefore does not approach one. Near the end of the branch the scalar fields follow a solution of the first-order equations, while the metric carries a nontrivial blackening factor. In this sense the solutions resemble the non-extremal D0 solution \eqref{eq:D0twodimensional}, whose scalars take their supersymmetric values while the metric is blackened. The geometry itself does not become supersymmetric.

The energy, by contrast, approaches its supersymmetric value. With the normalisation $E_{\rm BPS}=0$, the Smarr relation \eqref{eq:Smarr} gives $E-E_{\rm BPS}=\frac{9}{14}TS+\frac{5}{14}\Pi\mu$, and both terms vanish at the end of the branch. Numerically, $\widehat E/\widehat E|_{\mu=0}\propto(4\pi-\widehat\mu)^{3/5}$ and is dominated by the entropy term; at the reference point it equals $4.2\times10^{-4}$. This is compatible with a nontrivial blackening factor because the energy is fixed by the $w^7$ term of $1-b$ near the boundary, and this term vanishes in the limit. At $w\simeq0.1$, for instance, $(1-b)/w^7$ decreases from $0.084$ at $q_2^{\mathcal{H}}=0.52$ to $5.3\times10^{-3}$ at $q_2^{\mathcal{H}}=0.28$ and $4.6\times10^{-4}$ at $q_2^{\mathcal{H}}=0.157$, while $b(0.8)$ converges. The region in which $1-b$ behaves as $w^7$ shrinks towards the boundary as $a^3_0$ grows. The solutions therefore approach supersymmetry in their asymptotic charges, but not uniformly in the bulk.

\begin{table}[t]
\centering
\small
\setlength{\tabcolsep}{4pt}
\begin{tabular}{cccccccc}
\hline
$\widehat\mu$ & $q_2^{\mathcal{H}}$ & $a^3_0$ & $y_{\cal H}$ & $\delta_\Pi$ & $\delta_E$ & $\max|v-v_{\rm BPS}|$ & $\max|b_A/b_Z-1|$
\\
\hline
12.52658 & 0.5229 & 3.14 & $2.1\times10^{2}$ & $2.5\times10^{-3}$ & $1.1\times10^{-2}$ & $2.2\times10^{-3}$ & $3.6\times10^{-3}$\\
12.56596 & 0.2829 & 20.0 & $2.1\times10^{4}$ & $2.4\times10^{-5}$ & $8.4\times10^{-5}$ & $2.3\times10^{-5}$ & $3.9\times10^{-5}$\\
12.56636 & 0.1606 & 110 & $1.4\times10^{6}$ & $3.5\times10^{-7}$ & $1.2\times10^{-6}$ & $3.3\times10^{-7}$ & $5.5\times10^{-7}$\\
\hline
\end{tabular}
\caption{Approach of the thermal branch to a supersymmetric flow. The fifth and sixth columns measure the supersymmetric relations \eqref{eq:susyphivev} and \eqref{eq:susymetricvev} through $\delta_\Pi\equiv-a^4_2/(2\bar\mu a^3_0)-1$ and $\delta_E\equiv a^1_7\big/\bigl(\tfrac{5}{19}-\tfrac{472}{3591}\bar\mu^2a^3_0\bigr)-1$. The seventh column gives the largest deviation of $v(\log\rho)$ from the BPS trajectory \eqref{eq:flowlogrho} through the same solution, and the last compares the blackening factors extracted from $A$ and from $Z$ in \eqref{eq:blackenedBPS}, for $w\leq0.99$. The first two rows use $N=300$ and the last $N=400$.}
\label{tab:susylimit}
\end{table}

It remains to identify the trajectory. Along the thermal solutions we find $\eta\simeq y/y_{\cal H}$, with $y_{\cal H}=e^{3v_{\cal H}}$, to better than two per cent for $10^2\lesssim y\leq y_{\cal H}$. Comparing with \eqref{eq:largeyC} gives $C=y_{\cal H}^{-1}$, and hence $\eta_{\cal H}\to1$; at the reference point $1-\eta_{\cal H}=7\times10^{-7}$. Since $y_{\cal H}$ grows without bound, $C\to0$, and at fixed $y$ the thermal $\eta$ vanishes. The limiting trajectory is therefore $\eta=0$, which is \eqref{eq:Pstar}. This is the opposite end of the family from the explicit solution \eqref{eq:Pexplicit}, which has $C=1$.

Close to the end of the branch the solutions thus have two regions. Where $y\ll y_{\cal H}$ they approach the flow \eqref{eq:Pstar} of the ${\bf 44}$ scalar alone, with a large expectation value $a^3_0$. Where $y$ is comparable with $y_{\cal H}$, $\eta$ is of order one and the solution enters the large-$y$ region \eqref{eq:largeyC} of the BMN flows, which is singular as $y\to\infty$; see the behaviour of \eqref{eq:Pexplicit} as $z\to z_0$. The horizon lies in this second region and moves deeper into it as $q_2^{\mathcal{H}}\to0$. Near the end of the branch we therefore find a thermal geometry whose scalar trajectory approaches a supersymmetric flow, with a horizon moving into the singular large-$y$ region of that flow. The limit is not uniform. The outer limiting trajectory \eqref{eq:Pstar} has $\phi=0$, whereas every thermal solution satisfies the BMN boundary conditions, and the blackening factor remains nontrivial at fixed $w$. The vanishing of $\tilde S$, $\tilde F$ and $\tilde\Pi$ is consistent with this picture, but we have not shown that the branch ends on a supersymmetric solution.

\subsection{Horizon geometry near the end of the branch}\label{sec:horizongeometry}

The eleven-dimensional uplift gives a more direct picture of what happens to the horizon. From \eqref{eq:11dmetric}, the proper radius of its $S^2$ factor is ${\cal R}_2(\zeta)=\rho_{\cal H}^{2/9}e^{v_{\cal H}}\zeta\,{\cal H}_{\cal H}(\zeta)^{-1/3}$. At $\zeta=1$, where the $S^5$ shrinks, this radius develops a pronounced neck relative to its maximum over the horizon: ${\cal R}_2(1)/\max_\zeta{\cal R}_2(\zeta)$ falls from $0.85$ at $\widehat\mu=10.472$ to $0.24$ at $12.524$, $0.050$ at $12.566$ and $0.012$ at the reference point. The $S^2$ already closes off at $\zeta=0$, so a second pinch at $\zeta=1$ would change the way the horizon closes. Such a pinch would be compatible with a merger with a ringoid branch of $S^3\times S^5$ horizon topology after reduction on the M-theory circle, or $S^1\times S^3\times S^5$ in eleven dimensions, as suggested in Ref.~\cite{Costa:2014wya}. Every finite solution constructed here still has $S^1\times S^8$ topology.

The results of section~\ref{sec:susylimit} fix the form of the neck. Near $\zeta=1$ it is convenient to introduce $\sigma=e^{3v_{\cal H}/2}\sqrt{1-\zeta^2}$. For $y_{\cal H}\gg1$ and $1-\zeta^2\ll1$, \eqref{eq:DeltaF} gives
\begin{equation}
 \Delta_{\cal H}\simeq e^{-2v_{\cal H}}\left(1+\sigma^2\right),\qquad
 {\cal H}_{\cal H}\simeq1+\frac{4y_{\cal H}^2\eta_{\cal H}}{1+\sigma^2}\,.
\end{equation}
The flux term in ${\cal H}_{\cal H}$ dominates for $\sigma^2\ll y_{\cal H}^2$. In units of the neck, the radii of the two spheres and the proper distance $s$ along the horizon, measured from $\zeta=1$, are then
\begin{equation}
 \frac{{\cal R}_2}{{\cal R}_2(1)}\simeq\left(1+\sigma^2\right)^{1/3},\qquad
 \frac{{\cal R}_5}{{\cal R}_2(1)}\simeq\frac{2\eta_{\cal H}^{1/2}\sigma}{(1+\sigma^2)^{1/6}},\qquad
 \frac{s}{{\cal R}_2(1)}\simeq2\eta_{\cal H}^{1/2}\int_0^\sigma\left(1+\sigma'^2\right)^{1/3}{\rm d}\sigma' .
 \label{eq:neckprofile}
\end{equation}
This profile depends only on $\eta_{\cal H}$, which tends to one. It is therefore universal near the end of the branch. Indeed, at fixed $s/{\cal R}_2(1)$ we find the slopes ${\rm d}{\cal R}_i/{\rm d}s$ to be independent of $q_2^{\mathcal{H}}$ to three significant figures for $q_2^{\mathcal{H}}\lesssim0.3$, even though ${\cal R}_2(1)$ decreases by a factor of 40 over this range; at $s=8{\cal R}_2(1)$ they are $0.107$ and $0.241$. For $1\ll\sigma\ll y_{\cal H}^{1/2}$, \eqref{eq:neckprofile} gives
\begin{equation}
 {\cal R}_2\propto{\cal R}_5\propto s^{2/5},\qquad \frac{{\cal R}_5}{{\cal R}_2}\to2 .
 \label{eq:neckscaling}
\end{equation}
The radius of the M-theory circle decreases as $s^{-1/5}$ in the same region. The upper limit corresponds to $s/{\cal R}_2(1)\sim y_{\cal H}^{5/6}$, where $\zeta$ departs appreciably from one; at the reference point this is of order $10^5$. The numerical solutions reproduce \eqref{eq:neckprofile} and \eqref{eq:neckscaling}, as shown in Fig.~\ref{fig:bps}: at $s=10^3{\cal R}_2(1)$ the local exponent ${\rm d}\log{\cal R}_i/{\rm d}\log s$ is $0.400$ for both spheres.

Since the neck is universal in units of ${\cal R}_2(1)$, which tends to zero at fixed temperature, the curvature at the pinch must diverge as ${\cal R}_2(1)^{-2}$. We have checked this directly. The bifurcation surface of the horizon is a multiply warped product of the $S^2$, the $S^5$ and the M-theory circle over the interval $0\leq\zeta\leq1$, so its Kretschmann scalar $K_{\cal H}$ follows in closed form from the warp factors in \eqref{eq:11dmetric}. At $\zeta=1$, $K_{\cal H}$ grows from $3.3\times10^2$ at $\widehat\mu=8.38$ to $4.1\times10^{15}$ at $q_2^{\mathcal{H}}=0.111$, the smallest value we have reached, while
\begin{equation}
 K_{\cal H}\,{\cal R}_2(1)^4\big|_{\zeta=1}\longrightarrow15.9167 ,
 \label{eq:Kretschmann}
\end{equation}
to six significant figures for all $q_2^{\mathcal{H}}\leq0.195$. The solutions therefore become singular at the end of the branch, as the black branes of Ref.~\cite{Dias:2017stenzel} do.

\begin{figure}[t]
\centering
\includegraphics[width=\textwidth]{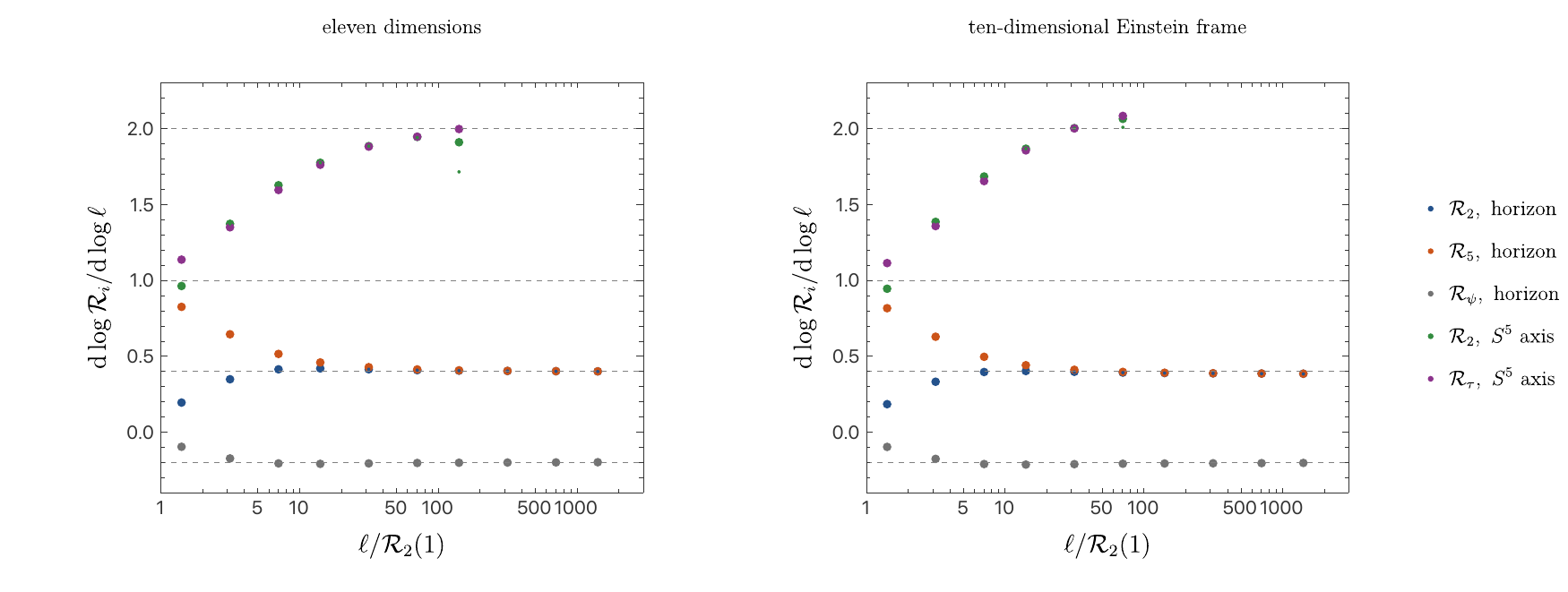}
\caption{Local exponents ${\rm d}\log{\cal R}_i/{\rm d}\log\ell$ along orbit-space geodesics leaving the pinch at $w=1$, $\zeta=1$, for the $N=400$ solution at $q_2^{\mathcal{H}}=0.161$. The upper points follow the $S^5$ axis outwards and the lower points follow the horizon. Small points show the $N=200$ solution at $q_2^{\mathcal{H}}=0.179$. Dashed lines mark the exponents $2$, $1$, $2/5$ and $-1/5$; a conical region would require exponent $1$.}
\label{fig:geodesics}
\end{figure}

A merger of the kind proposed in Ref.~\cite{Costa:2014wya} would be governed locally by a double-cone geometry \cite{Kol:2002xz,Emparan:2011ve}, in which all radii grow linearly with the distance from the pinch. In the profile \eqref{eq:neckscaling}, by contrast, the flux-dependent term in ${\cal H}$ dominates. To test the full geometry, and not only the horizon, we have followed geodesics of the orbit space of \eqref{eq:11dmetric} from the point $w=1$, $\zeta=1$. We used seven directions, from along the horizon to outwards along the $S^5$ axis. Along each geodesic we measured ${\cal R}_2$, ${\cal R}_5$, the radius ${\cal R}_\tau$ of the Euclidean time circle and the radius ${\cal R}_\psi$ of the M-theory circle as functions of the geodesic distance $\ell$. Figure~\ref{fig:geodesics} shows the local exponents along the two extreme directions, in eleven dimensions and in the ten-dimensional Einstein frame obtained by reducing on $\psi$.

Along the horizon, for $50\leq\ell/{\cal R}_2(1)\leq2000$, the exponents of ${\cal R}_2$ and ${\cal R}_5$ decrease from $0.413$ and $0.427$ to $0.400$ in eleven dimensions, and from $0.397$ and $0.410$ to $0.384$ in the Einstein frame. That of ${\cal R}_\psi$ is $-0.20$. Outwards along the $S^5$ axis, the exponents of ${\cal R}_2$ and ${\cal R}_\tau$ rise from $1.8$ at $\ell=20{\cal R}_2(1)$ to between $1.9$ and $2.0$ at $\ell=100{\cal R}_2(1)$; beyond this the geodesic leaves the near-horizon region. The solutions with $N=200$ at $q_2^{\mathcal{H}}=0.179$ and with $N=400$ at $q_2^{\mathcal{H}}=0.161$ give exponents which agree to $1.2\times10^{-3}$ along the horizon, and to $10^{-3}$ along the axis for $\ell\leq100{\cal R}_2(1)$. Geodesics leaving the pinch at intermediate angles bend back towards the horizon and show no scaling regime of their own. Within this range, therefore, no direction shows the linear growth required of a conical region, and the M-theory circle is not a spectator. This is strong evidence against a Ricci-flat double cone in the measured scaling regime. It does not exclude a merger with a ringoid branch through a different, flux-supported critical geometry, possibly at scales closer to the end point than we have resolved.

\section{Discussion}\label{sec:discussion}

We have followed the thermal branch of the BMN matrix model at strong coupling much further than was previously possible. The consistent truncation to two-dimensional $SO(9)$ gauged supergravity reduces the problem to ordinary differential equations. With a continuation in the horizon value $q_2^{\mathcal{H}}$ rather than in the mass, the branch can be followed to within $4\times10^{-7}$ of its end point in $\widehat\mu$, which power-law fits place at $\widehat\mu=4\pi$ with an estimated uncertainty below $10^{-10}$. The first-order transition inferred in Ref.~\cite{Costa:2014wya} does not survive: the black-hole free energy remains negative at the quoted temperature, and we trace the old crossing to the reconstruction of the free energy from a polynomial fit of the entropy. Instead, the free energy of the black holes approaches that of the confined phase continuously at the end of the branch, as $(\widehat\mu_c-\widehat\mu)^{8/5}$. If this branch remains the dominant one up to its end point, the transition coincides with the point at which the solutions become singular, as for mass-deformed M2 branes \cite{Dias:2017stenzel}.

The structure of the solutions near the end of the branch is also unexpected. As $\widehat\mu$ approaches $4\pi$, the asymptotic data satisfy the supersymmetric relations \eqref{eq:susyphivev} and \eqref{eq:susymetricvev} ever more accurately. In particular, the energy tends to its supersymmetric value, $E-E_{\rm BPS}\simeq TS\propto(4\pi-\widehat\mu)^{3/5}$. At every regular point on the branch, however, $E-E_{\rm BPS}>0$; its extrapolated vanishing occurs at a singular end point, not at a smooth finite-temperature black hole. The scalar fields follow a solution of the first-order equations from the boundary to the horizon. The metric, by contrast, differs from the supersymmetric one by the blackening factor in \eqref{eq:blackenedBPS}, which remains nontrivial at fixed radius. The scalar trajectory is not the explicit flow of section~\ref{sec:susy}. It approaches the other end of the one-parameter family of supersymmetric flows, the solution \eqref{eq:Pstar} with $\phi=0$, and the horizon moves into the singular large-$y$ region of these flows. Since the limiting trajectory has $\phi=0$ while the BMN boundary conditions do not change, this limit is not uniform. The vanishing of $\tilde S$, $\tilde F$ and $\tilde\Pi$ is consistent with this structure, but it remains to be shown whether the branch ends on a supersymmetric solution in any precise sense.

The pronounced neck which develops on the horizon has a correspondingly simple origin. In units of the neck its profile is universal, and the radii of both spheres grow as the $2/5$ power of the proper distance from the pinch. In this region the flux-dependent term in ${\cal H}$ dominates the uplift. Within the range we can resolve we find no direction along which the geometry is conical. This is strong evidence against a Ricci-flat double cone of the kind that controls the merger of vacuum black holes with black strings \cite{Kol:2002xz,Emparan:2011ve}. It does not exclude a merger with a ringoid branch through a different, flux-supported critical geometry, possibly at scales closer to the end point than we have reached.

Several questions remain. We have no analytic argument that fixes the end of the branch at $\widehat\mu=4\pi$ or the exponents in \eqref{eq:endexponents}, nor do we have the blackening function in closed form. Both might follow from the structure found here, perhaps together with the general solution of \eqref{eq:Pequation}, which we have not found. Since the transition coincides with a singularity of supergravity, its fate beyond the supergravity approximation is also open. It would also be important to reach this regime with the eleven-dimensional conformal-coordinate calculation of Appendix~\ref{app:conformal}. That calculation tests the reduction independently and can search for instabilities which break $SO(3)\times SO(6)$. Finally, the global phase structure requires comparison with other black-hole branches and with the confined backgrounds associated with the BMN vacua.

\acknowledgments

It is a pleasure to thank Juan Maldacena for initiating this project with me, and for insisting, with admirable persistence, that this material should be published. I am also grateful to him for reading an earlier draft. JES is supported by STFC grant number ST/X000664/1 and partially by Hughes Hall, Cambridge. I also thank Claude, an AI model developed by Anthropic, which was used to check the consistency of factors of $2$ and $\pi$ throughout the manuscript. This proved invaluable. My unwavering attachment to the mostly-plus signature means that every sign and factor imported from the literature must be rederived, and preferably by someone who does not get bored.

\appendix

\section{Covariant phase space derivation of the energy}\label{app:wald}
We now give an independent derivation of the charges in section~\ref{sec:holoenergy} using the covariant phase space formalism of Wald \cite{Wald:1999wa}. Since the solutions we seek to construct have a Killing vector field $\xi$, with components $\xi^a$, that is timelike near the boundary, we define an energy with respect to such a Killing field. For the moment let us take $\xi$ as an arbitrary diffeomorphism generator, and specialise to a Killing vector field near the end.

Varying the bulk action \eqref{eq:actionbmn} generates the following boundary terms:
\begin{subequations}\label{eqs:boundaryterms}
\begin{align}
&2\kappa_2^2\theta^g(g,\delta g)=\star\left\{\frac{1}{4}\rho\left[\nabla^b \delta g_{ab}-\nabla_a (g^{bc}\delta g_{bc})\right]\mathrm{d}x^a-\frac{1}{4}\left[\delta g_{ab}\nabla^b\rho-\nabla_a \rho(g^{bc}\delta g_{bc})\right]\mathrm{d}x^a\right\}
\\
& 2\kappa_2^2\theta^v(v,\delta v)=-\frac{9}{4}\rho \star \mathrm{d}v \wedge\delta v
\\
& 2\kappa_2^2\theta^\phi(\phi,\delta \phi)=-e^{6v} \rho^{1/3}\star \mathrm{d}\phi\wedge \delta \phi\,.
\end{align}
\end{subequations}
Next, we evaluate the above on variations $\{\delta v,\delta \phi,\delta g\}$ associated to diffeomorphisms generated by $\xi$ and define
\begin{equation}
\Theta^g(g,\pounds_{\xi}g)\equiv \theta^g(g,\pounds_{\xi}g)\,,
\end{equation}
and similarly for $v$ and $\phi$, where we set $\delta v=\pounds_{\xi}v = \xi^a\nabla_a v$, $\delta \phi =\pounds_{\xi}\phi= \xi^a\nabla_a \phi$ and $\delta g_{ab}=(\pounds_{\xi}g)_{ab}=2\nabla_{(a}\xi_{b)}$. A few lines of somewhat tedious but straightforward algebra show
\begin{multline}
2\kappa_2^2\Theta^g(g,\pounds_{\xi} g)=\star\Bigg\{\frac{1}{2}\left[\nabla^b(\rho \nabla_{[b}\xi_{a]})-2 \nabla^b(\nabla_{[b}\rho\xi_{a]})\right]\mathrm{d}x^a
\\
+\frac{\rho}{2}\xi^b\left(R_{ab}-\frac{1}{\rho}\nabla_a\nabla_b\rho+g_{ab}\frac{\Box \rho}{\rho}\right)\mathrm{d}x^a\Bigg\}\,.
\end{multline}
Finally, a conserved current can be constructed from the above by considering
\begin{equation}
J=\Theta^g(g,\pounds_{\xi}g)+\Theta^v(v,\pounds_{\xi}v)+\Theta^\phi(\phi,\pounds_{\xi}\phi)-i_{\xi}\left.\mathcal{L}\right|_{\rm on-shell}\,,
\end{equation}
where $\mathcal{L}$ is the Lagrangian density (seen as a two-form) that follows from our action (\ref{eq:actionbmn}). Furthermore, $i_{\xi}\omega_p$ is the interior product of a general $p$-form $\omega_p$ under $\xi$. To evaluate the on-shell action we use Eq.~(\ref{eq:rho}), to find
\begin{equation}
2\kappa_2^2\left.i_{\xi}\mathcal{L}\right|_{\rm on-shell}=\star\left\{\left(-\frac{1}{3}e^{6v}\rho^{1/3}\nabla_b \phi\nabla^b\phi+\rho \frac{\partial V}{\partial \rho}-V\right)\xi_a\,\mathrm{d}x^a\right\}\,
\end{equation}
which finally yields
\begin{equation}
2\kappa_2^2 J = -\frac{1}{4} \mathrm{d}\left[\rho\,\star \, \mathrm{d}\xi^\flat-2 \star(\mathrm{d}\rho\wedge \xi^\flat)\right]\,,
\end{equation}
where $\xi^\flat=\xi_a\,{\rm d}x^a$, and we used $\star^2=1$ on one-forms and Eq.~(\ref{eq:mog}). On zero-forms and two-forms our Lorentzian Hodge dual instead satisfies $\star^2=-1$.

Clearly, $\mathrm{d}J=0$, and thus we can write $ J=\mathrm{d}Q_{\xi}$ with
\begin{equation}
Q_{\xi}=-\frac{1}{8\kappa_2^2}\left[\rho\,\star \, \mathrm{d}\xi^\flat-2 \star(\mathrm{d}\rho\wedge \xi^\flat)\right].
\label{eq:Waldcharge}
\end{equation}

For the static metric \eqref{eq:metricAnsatz} and $\xi=\partial_t$, this gives
\begin{equation}
 Q_{\partial_t}=\frac{e^{A-Z}}{4\kappa_2^2}\left(\rho A'-\rho'\right).
 \label{eq:staticWaldcharge}
\end{equation}
At a regular non-extremal horizon, $e^{A-Z}A'$ tends to the positive surface gravity, while $e^{A-Z}\rho'$ vanishes. This fixes the sign of the horizon charge with our orientation ${\rm vol}_2=e^{A+Z}{\rm d}t\wedge{\rm d}r$.

We are interested in variations of these charges, so that one can define an energy via the first law. The covariant phase space charge variation is \cite{Wald:1999wa}
\begin{equation}
\omega_{\xi}=\delta Q_{\xi}-Q_{\delta\xi}-i_{\xi}\theta,
\qquad \theta=\theta^g+\theta^v+\theta^\phi.
\end{equation}
Here $\delta$ acts on all fields $\{g_{ab},\rho,v,\phi\}$; there is no separate $\theta^\rho$ because the bulk action contains no derivatives of $\rho$. We keep $\xi=\partial_t$ fixed, so that $Q_{\delta\xi}=0$. For a Killing vector and variations satisfying the linearised equations, $\mathrm{d}\omega_{\xi}=0$, so that we can equate its value on the horizon to its value near the boundary.

With the orientation above, $\omega_{\xi}$ on the horizon yields the $T\,\mathrm{d}S$ term in the first law, where $T$ is the temperature and $S$ the black hole entropy. Near the boundary, the same charge variation takes the form $\mathrm{d}E-\Pi\,\mathrm{d}\mu$, where $\Pi$ is the thermodynamic response to the mass deformation. Equating the two gives
\begin{equation}
\mathrm{d}E-\Pi \mathrm{d}\mu=T \mathrm{d}S\,.
\end{equation}

It is a simple exercise to compute the Hawking temperature of the black hole, for which we find
\begin{equation}
T=\frac{7 g^{7/2}r_+^{5/2}}{4\pi}\,.
\end{equation}
Evaluating $\omega_{\xi}$ on the horizon, and using the above expression for the temperature, allows us to read off the entropy, which turns out to be
\begin{equation}
S=\frac{\pi g^{9/2}r_+^{9/2}}{2\kappa_2^2}q_2(1)^{9/2}=\frac{\pi}{2\kappa_2^2}\rho(r_+)\,.
\end{equation}
The Einstein term in our action is $\rho R/(8\kappa_2^2)$. Identifying $\kappa_2^2=2\pi G_2$, where $G_2$ is the two-dimensional Newton's constant, puts this term in the standard form $\rho R/(16\pi G_2)$. The entropy can then be written as
\begin{equation}
S=\frac{\rho(r_+)}{4 G_2}\,.
\end{equation}

At infinity we use the asymptotic expansion \eqref{eq:asymptoticqi} and the mass convention \eqref{eq:mudef}.

By evaluating $\omega_{\xi}$ near the boundary, we find $\mathrm{d}E-\Pi \mathrm{d}\mu$, which can be integrated to determine $E$ and $\Pi$ up to an arbitrary integration function $\mathfrak{f}(\mu)$
\begin{subequations}
\begin{align}
&E=\frac{g^{8}r_+^7}{840}(945-3591 a^1_7-2248 \bar{\mu}^2a^3_0-888 \bar{\mu} a^4_2)+\mathfrak{f}(\mu)\,,
\\
& \Pi =-\frac{1}{2}\,g^{9/2} r_+^{9/2}(2\bar{\mu}a^3_0+a^4_2)+\frac{\mathrm{d}\mathfrak{f}(\mu)}{\mathrm{d}\mu}\,,
\end{align}
\label{eqs:thermo}
\end{subequations}
where we used the BMN source condition \eqref{eq:susybc} and, as in \eqref{eqs:holothermo}, units $2\kappa_2^2=1$. The function $\mathfrak f(\mu)$ reflects the freedom to shift the energy by a local function of the source while preserving the first law. Matching to the counterterm prescription generated by $W$ in section~\ref{sec:renormalisation} sets $\mathfrak f=0$ and reproduces \eqref{eqs:holothermo}. The vanishing charges \eqref{eq:BPScharges} provide a check of this choice.

\section{Supersymmetric flow and boundary prescription}\label{app:susyboundary}

\subsection{Fermion conventions and Killing spinors}\label{app:killingspinors}

We spell out the fermion conventions needed to reproduce the flow. There are sign differences between the formulas as printed in \cite{Ortiz:2012ib} and the mutually consistent set used here. In the notation of that reference, we take the $\Gamma^{abc}b^{abc}$ contribution to $B_{IJ}$ with a plus sign, and the corresponding contribution to $\widetilde D_{IJ}$ with coefficient $-20/81$. Before converting the spacetime Clifford algebra to mostly-plus signature, the axion derivative terms in $\delta\psi_\mu$ and $\delta\chi^a$ have coefficients $+1/24$ and $+i/6$, respectively, and the matter-fermion shift is $C^a-\widetilde C^a\gamma^3$. The reduced equations below fix these conventions without reference to a particular matrix representation.\footnote{The correction of the quartic term in $b^a$ from $1/144$ to $-1/288$ is recorded in \cite{Anabalon:2013zka}. This term vanishes in our one-axion sector. In the same reference, the factor multiplying $(\rho y^a+8\prod_{b\ne a}\phi^b)^2$ in equation~(2.11) should be $X_a^2$. That term also vanishes in the sector used here.}

Let $\gamma^{\hat t}$ and $\gamma^{\hat r}$ obey the mostly-plus Clifford algebra, and define
\begin{equation}
 H=\gamma^{\hat r},\qquad
 \gamma_*=-\gamma^{\hat t}\gamma^{\hat r},\qquad
 {\cal K}=\Gamma^{129},\qquad M=\gamma_*{\cal K}.
 \label{eq:spinoralgebra}
\end{equation}
The internal matrices satisfy $\{\Gamma^a,\Gamma^b\}=2\delta^{ab}$. Consequently $H^2=1$, $M^2=-1$, $\{H,M\}=0$, and ${\cal K}$ commutes with both $H$ and $M$. A dot denotes differentiation with respect to proper radial distance, $\dot f=e^{-Z}f'$. It is useful to introduce
\begin{align}
 \mathfrak b&=\frac{3g}{4}\rho^{-2/9}(e^{-2v}+2e^v),&
 \mathfrak d&=\frac{3g}{2}\rho^{-5/9}e^v\phi,\nonumber\\
 \mathfrak c&=-\frac{g}{3}\rho^{-2/9}(e^{-2v}-e^v),&
 {\cal Q}&=\sqrt{\mathfrak b^2+\mathfrak d^2},\qquad
 y=\rho^{-1/3}e^{3v}\dot\phi.
 \label{eq:reducedshifts}
\end{align}
Here $y$ is a shorthand for the axion derivative, unrelated to the auxiliary fields set to zero in the truncation. We take $g>0$ and $\rho>0$, so $\mathfrak b>0$.

The nonzero gravitino and dilatino shifts are
\begin{equation}
 \mathsf A=\frac79\mathfrak b\,\mathbbm{1},\qquad
 \widetilde{\mathsf A}=-\frac49\mathfrak d\,{\cal K},\qquad
 \mathsf B=\mathfrak d\,{\cal K},\qquad
 \widetilde{\mathsf B}=\mathfrak b\,\mathbbm{1}.
 \label{eq:reducedABshifts}
\end{equation}
For example, these satisfy $\mathsf A-\widetilde{\mathsf B}-\rho\partial_\rho\widetilde{\mathsf B}=0$ and $\widetilde{\mathsf A}+\mathsf B+\rho\partial_\rho \mathsf B=0$, with the scalar fields held fixed in the derivatives. The latter identity would fail with the sign of $\mathsf B$ printed in equation~(4.17) of \cite{Ortiz:2012ib}. Likewise $\widetilde{\mathsf D}=-20\mathfrak d\,{\cal K}/81$ satisfies $\widetilde{\mathsf D}+\rho\partial_\rho\widetilde{\mathsf A}=0$.

After removing invertible gamma-matrix factors, the dilatino and the independent matter-fermion equations reduce to
\begin{align}
 0&=\left[-\frac{\dot\rho}{2\rho}H+\mathfrak b+\mathfrak d M\right]\epsilon,
 \label{eq:reduceddilatino}\\
 0&=\left[\frac{\dot v}{2}H-\frac{y}{3}HM
       +\mathfrak c+\frac{2\mathfrak d}{9}M\right]\epsilon.
 \label{eq:reducedmatter}
\end{align}
The matter equations in the other six internal directions are proportional to \eqref{eq:reducedmatter}. On the branch with increasing $\rho$, the first equation implies
\begin{equation}
 \dot\rho=2\rho{\cal Q},\qquad
 {\cal P}\epsilon=\epsilon,\qquad
 {\cal P}=\frac{\mathfrak b H+\mathfrak d HM}{{\cal Q}},
 \qquad {\cal P}^2=1.
 \label{eq:BPSprojector}
\end{equation}
Using this projector in \eqref{eq:reducedmatter} and separating the coefficients of $\epsilon$ and $M\epsilon$ gives
\begin{equation}
 \dot v=-\frac{2}{{\cal Q}}
    \left(\mathfrak b\mathfrak c+\frac{2}{9}\mathfrak d^2\right),
 \qquad
 y=\frac{3\mathfrak d\mathfrak c-\frac23\mathfrak b\mathfrak d}{{\cal Q}}.
 \label{eq:BPSmatterflow}
\end{equation}
These are precisely the $\rho$, $v$ and $\phi$ equations in \eqref{eq:firstordersystem}.

The gravitino equations complete the check. Define $\theta$ by
$\cos\theta=\mathfrak b/{\cal Q}$ and $\sin\theta=\mathfrak d/{\cal Q}$, with $\theta=0$ at $\phi=0$. Requiring the time component of the gravitino equation to preserve \eqref{eq:BPSprojector} gives
\begin{equation}
 \dot A=\frac{2(7\mathfrak b^2+4\mathfrak d^2)}{9{\cal Q}}.
 \label{eq:BPSmetricfromspinor}
\end{equation}
The remaining spinor equations are
\begin{equation}
 \dot\epsilon=\left(\frac{\dot A}{2}-\frac{\dot\theta}{2}M\right)\epsilon,
 \qquad
 \partial_t\epsilon=\Omega{\cal K}\epsilon,
 \qquad
 \Omega=e^A\left(\frac{y}{6}+\frac{\mathfrak b\mathfrak d}{3{\cal Q}}\right).
 \label{eq:spinordifferential}
\end{equation}
Substitution of \eqref{eq:BPSprojector}, \eqref{eq:BPSmatterflow} and \eqref{eq:BPSmetricfromspinor} gives
\begin{equation}
 \dot\theta=\frac23\left(y-\frac{\mathfrak b\mathfrak d}{{\cal Q}}\right),
 \qquad \dot\Omega=0.
 \label{eq:spinorintegrability}
\end{equation}
Thus the Killing spinor can be written locally as
\begin{equation}
 \epsilon(t,r)=e^{A(r)/2}\,e^{-\theta(r)M/2}\,
 e^{t\Omega{\cal K}}\epsilon_0,
 \qquad H\epsilon_0=\epsilon_0,
 \label{eq:explicitKillingspinor}
\end{equation}
where $\epsilon_0$ is constant. The time dependence is essential when the mass deformation is nonzero. Finally,
\begin{equation}
 \rho{\cal Q}=W,\qquad
 \frac{2(7\mathfrak b^2+4\mathfrak d^2)}{9{\cal Q}}
 =2\partial_\rho W,
 \label{eq:spinorWidentification}
\end{equation}
which identifies the entire system with \eqref{eq:HJflow} independently of the bosonic Hamilton--Jacobi construction.

\subsection{The bosonic boundary variation}\label{app:BPSboundaryvariation}

There is an exact identity for the radial action that makes the role of $W$ explicit. Define the departures from the BPS flow by
\begin{align}
 {\cal B}_\rho&=\dot\rho-2W,&
 {\cal B}_A&=\dot A-2\partial_\rho W,\nonumber\\
 {\cal B}_v&=\dot v+\frac{4}{9\rho}\partial_v W,&
 {\cal B}_\phi&=\dot\phi+e^{-6v}\rho^{-1/3}\partial_\phi W.
 \label{eq:BPSresiduals}
\end{align}
Writing $S_{\rm red}=(2\kappa_2^2)^{-1}\int {\rm d}r\,{\rm d}t\,{\cal L}_{\rm red}$, the Hamilton--Jacobi identity gives
\begin{equation}
 {\cal L}_{\rm red}
 =\partial_r(e^A W)+e^{A+Z}
 \left[\frac12{\cal B}_A{\cal B}_\rho
 -\frac98\rho{\cal B}_v^2
 -\frac12e^{6v}\rho^{1/3}{\cal B}_\phi^2\right].
 \label{eq:BPSactioncompletion}
\end{equation}
This identity is valid off shell in the static bosonic sector. It is a completion in products of flow residuals, not a positivity statement for the Lorentzian gravitational action. The outer boundary contribution is cancelled by \eqref{eq:susycounterterm}; any inner boundary must be treated separately.

More directly, the canonical momenta obey
\begin{align}
 \pi_A-e^AW&=\frac12e^A{\cal B}_\rho,&
 \pi_\rho-e^A\partial_\rho W&=\frac12e^A{\cal B}_A,\nonumber\\
 \pi_v-e^A\partial_v W&=-\frac94e^A\rho{\cal B}_v,&
 \pi_\phi-e^A\partial_\phi W&=-e^Ae^{6v}\rho^{1/3}{\cal B}_\phi.
 \label{eq:BPSmomentumidentity}
\end{align}
All four outer boundary momenta therefore vanish on the BPS flow before the Legendre transform. The Legendre term exchanges the scalar boundary coordinate and momentum; it does not alter this conclusion on the branch with vanishing independent scalar source.

Equations \eqref{eq:BPSactioncompletion}--\eqref{eq:BPSmomentumidentity} establish the boundary variation on the BPS flow in the bosonic sector.

\subsection{Scaling and the finite counterterm class}\label{app:countertermscaling}

The dilaton homogeneity used in \eqref{eq:countertermclass} can be tied to a scaling of the classical bulk action. At fixed $g$ and $\kappa_2$, the constant transformation
\begin{equation}
 \rho\longrightarrow s\rho,\qquad
 \phi\longrightarrow s^{1/3}\phi,\qquad
 g_{ab}\longrightarrow s^{4/9}g_{ab},\qquad
 v\longrightarrow v,\qquad s>0,
 \label{eq:dilatonscaling}
\end{equation}
multiplies both $S_{\rm bulk}$ and $S_{\rm GH}$ by $s$. Thus it is a scaling covariance of the classical action and its equations, rather than an invariance of the action at fixed normalisation. We require the counterterm prescription to respect this covariance. Since $\sqrt{-\gamma}\longrightarrow s^{2/9}\sqrt{-\gamma}$, a zero-derivative counterterm must satisfy
\begin{equation}
 U(s\rho,v,s^{1/3}\phi)=s^{7/9}U(\rho,v,\phi).
 \label{eq:countertermhomogeneity}
\end{equation}
Its general form is \eqref{eq:countertermclass}, up to coefficients depending on fixed parameters. Analyticity in $v$ and $x$ near the undeformed solution then gives the monomial classification in \eqref{eq:countertermweight}. The only finite term in this class is $vx^2$.

The derivative expansion has a separate weight count. Under \eqref{eq:time-dependent-falloffs}, the density of \eqref{eq:time-dependent-counterterm} is $O(r^2)$, while the next pure-dilaton term, with four boundary derivatives, is $O(r^{-3})$. The scalar kinetic densities $\sqrt{-g}\,\rho(Dv)^2$ and $\sqrt{-g}\,e^{6v}\rho^{1/3}(D\phi)^2$ are radially integrable. At two-derivative order, a potentially finite term linear in $v$ could be $\sqrt{-\gamma}\,v(D\rho)^2/W_0$ or, after integration by parts, a term involving $D\rho\cdot Dv$. To test it, extend \eqref{eq:HJtwo-derivative-ansatz} by $v f_1(\rho)(D\rho)^2+h_0(\rho)D\rho\cdot Dv$. At $v=0$, the coefficient of $\Box_\gamma v$ in the Hamilton--Jacobi constraint is $-2W_0h_0$, so $h_0=0$. The remaining coefficient of $D\rho\cdot Dv$ is $-4W_0f_1$, so $f_1=0$. Thus the analytic Hamilton--Jacobi solution produces no such finite derivative term.

Both requirements matter. Without the scaling condition, the local densities $\sqrt{-\gamma}\,g\rho^{-7/9}$ and $\sqrt{-\gamma}\,g\rho^{-1/3}\phi^2$ have finite limits and allow additional source-only shifts. Without analyticity, even a scaling-compatible term such as $\sqrt{-\gamma}\,g\rho^{7/9}|x|^{14/5}$ is finite. These examples show why finiteness alone does not fix the scheme. Boundary derivative terms vanish, together with their first variations about constant sources, for the stationary observables considered here. Extrinsic-curvature or radial-momentum additions are not part of the intrinsic counterterm class with the specified variational principle.

A remaining source-only shift in a wider scheme changes $E$ by $f(\mu)$ and $\Pi$ by $f'(\mu)$. Requiring $E_{\rm BPS}(\mu)=0$ for every mass on the branch fixes this freedom on that branch. Counterterms that change the scalar boundary condition instead change the deformation being studied and must be distinguished from this choice of energy normalisation.

\subsection{The source condition from the fermion equation}\label{app:BPSsource}

The source relation can be obtained before introducing counterterms. Expanding the scalar equation \eqref{eq:BPSmatterflow} about the D0 boundary gives
\begin{equation}
 {\cal B}_v
 =\frac{w^5}{r}\left(\frac49\bar\mu^2-3a^3_3\right)
 +o\!\left(\frac{w^5}{r}\right).
 \label{eq:BPSsourceresidual}
\end{equation}
Here the lower-order terms cancel after the asymptotic field equations are used. Since \eqref{eq:BPSmatterflow} follows from the matter-fermion variation, the supersymmetric locus of the bosonic boundary data is therefore
$a^3_3=4\bar\mu^2/27$, independently of a counterterm choice.

The boundary variation determines how this locus is expressed as a source in alternate quantisation. Because $\delta v=w^2\delta a^3_0+\ldots$, the finite momentum conjugate to $a^3_0$ is obtained from $w^2\Pi_v^{\rm ren}$. Equation~\eqref{eq:BPSmomentumidentity} yields
\begin{equation}
 \lim_{w\to0}\frac{w^2\Pi_v^{\rm ren}}{g^8r_+^7}
 =\frac{27}{4}a^3_3-\bar\mu^2.
 \label{eq:sourcefromfermion}
\end{equation}
Using $U_c$ instead of $W$ shifts the right-hand side by $-(c-1)\bar\mu^2$. Thus the coefficient of the finite mixed term in the analytic, scaling-covariant class is fixed by matching the independently determined supersymmetric locus to the BMN condition of no additional ${\bf44}$ source.

This determines the source condition on the stationary bosonic backgrounds used in this paper.

\section{Integration of the supersymmetric flow}\label{app:flowintegration}

In this appendix we integrate the explicit solution \eqref{eq:Pexplicit} of the first-order flow and derive its asymptotic behaviour, which is summarised in section~\ref{sec:susy}. With $\dot v=P$, this solution gives
\begin{equation}
\dot{v}=1+\frac{5}{3}e^{3v}-\frac{4}{3}e^{6v}-\frac{4}{3}e^{9v}\,.
\end{equation}

This last equation can be readily integrated
\begin{equation}
\frac{16 e^{15 v}e^{-15(z-z_0)}(3+2 e^{3v})}{(e^{3v}-1)(1+2e^{3v})^5}-1=0\,,
\end{equation}
where $z_0$ is an integration constant whose physical meaning we will clarify later. For now we note that we can readily expand the above equation near $z=z_0$ to find
\begin{multline}
v\approx -\frac{1}{9}\log\left[12(z-z_0)\right]-\frac{1}{2^{4/3}3^{2/3}}(z-z_0)^{1/3}+\frac{1}{ 2^{2/3} 3^{1/3}}\frac{47}{40}(z-z_0)^{2/3}
\\
+\frac{73}{240}(z-z_0)+\mathcal{O}[(z-z_0)^{4/3}]
\end{multline}
while at large $z$ we find
\begin{equation}
v\approx \frac{80}{729} e^{-15 (z-z_0)}+\frac{30080 }{531441}e^{-30 (z-z_0)}+\frac{14766080}{387420489}e^{-45(z-z_0)}+\mathcal{O}\left(e^{-60 z}\right)
\end{equation}

It is a simple exercise to write the remaining variables in terms of $v$. Indeed, we find
\begin{subequations}
\begin{align}
&\dot{A}=\frac{1}{6} \left(1+2 e^{3 v}\right)^2 \left(3+4 e^{3 v}\right)
\\
& \frac{\dot{\rho}}{\rho}=\frac{3}{2}e^{3 v}(1+2 e^{3v})^2
\\
& Z= \log \left(1+2 e^{3 v}\right)+\frac{2}{9} \log \rho+\frac{7 v}{2}
\\
& \phi = \frac{1}{2} e^{-3 v} \sqrt{4 e^{9 v}-3 e^{3 v}-1} \rho^{1/3}
\end{align}
\end{subequations}

At large $z$ these yield
\begin{subequations}
\begin{align}
&A=\frac{21 z}{2}+A_0-\frac{320}{729} e^{-15 (z-z_0)}-\frac{1280}{6561} e^{-30 (z-z_0)}+\mathcal{O}\left(e^{-45 z}\right)
\\
&\rho = \rho _0 e^{\frac{27}{2} z} \left[1-\frac{56}{81} e^{-15 (z-z_0)}-\frac{544 }{6561}e^{-30 (z-z_0)}+\mathcal{O}\left(e^{-45 z}\right)\right]
\\
&Z=3 z+\log 3+\frac{2}{9} \log\rho _0+\frac{328}{729} e^{-15 (z-z_0)}+\frac{44608 }{177147}e^{-30 (z-z_0)}+\mathcal{O}\left(e^{-45 z}\right)
\\
&\phi = \frac{2 \sqrt{5} \rho_0^{1/3}e^{\frac{9 z_0}{2}} }{3 \sqrt{3}}e^{-3 (z-z_0)}+\mathcal{O}\left(e^{-48 z}\right)
\end{align}
\end{subequations}
where $A_0$ and $\rho_0$ are integration constants. Define
\begin{equation}
r=\frac{1}{3 g}\log\frac{g R}{\rho _0^{2/9}}+\sum_{i=5}^{+\infty}\frac{a_i}{R^i}\,
\end{equation}
and choose the coefficients $a_i$ so that to all orders in the $1/R$ expansion we have
\begin{equation}
g_{RR}=1\,.
\end{equation}

In such coordinates we find
\begin{equation}
g_{tt}=-\frac{g^7 R^7}{\rho _0^{14/9}}+\mathcal{O}(R^3)\,
\end{equation}
which shows that the metric has the expected behaviour for approaching a D0 brane asymptotically at large $R$. For the scalar fields we find
\begin{subequations}
\begin{equation}
\phi = g^{3/2}\frac{m_{\rm flow}}{R}+\mathcal{O}(R^{-6})
\end{equation}
and
\begin{equation}
v = \frac{4 m_{\rm flow}^2}{27 R^5}+\mathcal{O}(R^{-10})
\label{eq:BPSvasymptotic}
\end{equation}
so long as we make the identification
\begin{equation}
m_{\rm flow} = \frac{2 }{3}\sqrt{\frac{5}{3}} \frac{\rho _0^{5/9} }{g^{5/2}}e^{\frac{15 z_0}{2}}\,,
\end{equation}
\end{subequations}
The coefficient $m_{\rm flow}$ is expressed in the radial gauge used to integrate the flow. The coordinate $R$ agrees asymptotically with the static radial coordinate, so $m_\phi=g^{3/2}m_{\rm flow}$. After normalising the boundary time, the BMN mass is $\mu=4g^{7/2}m_{\rm flow}$ by \eqref{eq:massuplift}. Thus $z_0$ controls the mass deformation of the BMN model.

We now investigate the behaviour of the metric near $z=z_0$. In particular, just like we did for large $r$ (or $z$) we can change to a new coordinate $\chi$
\begin{equation}
r=\frac{z_0}{g}\left(1+\sum_{i=1}^{+\infty}b_i \chi^{\frac{12+6i}{7}}\right)
\end{equation}
and adjust the coefficients $b_i$ so that $g_{\chi\chi}=1$. In such coordinates we find that the two dimensional metric takes the following simple form
\begin{equation}
\mathrm{d}s^2\approx -\alpha^2\chi^{8/7}\mathrm{d}t^2+\mathrm{d}\chi^2+\mathcal{O}(\chi^2)
\end{equation}
where $\alpha$ is a known constant. Additionally, we find
\begin{equation}
\rho =\beta \chi^{9/7}+\mathcal{O}(\chi^{15/7})\qquad \phi =\gamma+\mathcal{O}(\chi^{18/7})\,,\qquad\text{and}\qquad v=-\frac{2}{7}\log \chi+\delta+\mathcal{O}(\chi^{6/7})\,,
\end{equation}
where $\beta$, $\gamma$ and $\delta$ are known constants.

\section{Conformal coordinates}\label{app:conformal}
In this appendix we improve the eleven-dimensional construction of Ref.~\cite{Costa:2014wya}, using conformal coordinates on the two-dimensional orbit space in place of the Einstein--DeTurck formulation. The collocation Jacobian is severely ill-conditioned in both formulations, but for different reasons. In the DeTurck system a handful of near-null modes are concentrated in the off-diagonal metric component, which enters the ansatz with a high power of the radial coordinate and obeys a Robin condition at the boundary. The associated small singular values decrease with resolution: the condition number is of order $10^{14}$ on a $16\times16$ grid and $10^{21}$ on a $30\times60$ grid. This limits Newton's method to modest grids, even at higher arithmetic precision. With the conversion \eqref{eq:CGPSmassconversion}, the saved free-energy curve of Ref.~\cite{Costa:2014wya} departs from ours already at small mass: at $\mu/T\simeq0.9$ the difference in $\tilde F$ is about $1.2\times10^{-3}$. As we discuss at the end of this appendix, most of this difference is removed by a rescaling of the horizontal axis, after which the two curves differ by $10^{-3}$ only beyond $\mu/T\simeq2.9$.

In conformal gauge the near-null modes instead arise from the physical fall-off of the $T$--$Z$ metric component. They do not prevent convergence: exact dense factorisation followed by iterative refinement in extended precision takes the calculation to round-off. The linearised operator is also the physical one, so we can follow its spectrum along the nonlinear branch and search directly for zero modes signalling a bifurcation to nearby solutions with less symmetry. We find no eigenvalue crossing zero among modes preserving $SO(3)\times SO(6)$ on the branch up to $\widehat\mu\simeq11.22$, providing perturbative evidence that no such nearby branch bifurcates from these solutions in this range.

We assume stationarity, invariance under translations along the M-theory circle, and an $SO(3)\times SO(6)$ symmetry.  The most general fields consistent with these symmetries, after fixing the residual coordinate freedom on the two-dimensional orbit space in conformal gauge, can be written as
\begin{align}
\frac{{\rm d}s_{11}^2}{L^2}
={}&
 -q_1(x,y)\,\frac{1-y^7}{y^7}\,y_+^7\,{\rm d}T^2 +\frac{y^7}{y_+^7}\,q_3(x,y) \left[ {\rm d}Z +q_6(x,y)\frac{1-y^7}{y^7}y_+^7\,{\rm d}T\right]^2
\nonumber
\\
&+\frac{y_+^2}{y^2}\bigg\{ q_2(x,y)
 \left[\frac{{\rm d}y^2}{(1-y^7)y^2} +\frac{4\,{\rm d}x^2}{2-x^2} \right]\nonumber
 \\
&\hspace{2.0cm}
+x^2(2-x^2)\,q_4(x,y)\,{\rm d}\Omega_2^2 +(1-x^2)^2q_5(x,y)\,{\rm d}\Omega_5^2
 \bigg\},
\label{eq:11d_ansatz}
\end{align}
Here ${\rm d}\Omega_2^2$ and ${\rm d}\Omega_5^2$ are the unit-radius round metrics on $S^2$ and $S^5$, respectively. Using the same $(y_1,\varphi_1)$ coordinates and orientation as in \eqref{eq:11dthreeform}, the three-form potential is
\begin{equation}
 C_{(3)}=L^3\left[\frac{y_+^3(1-y^7)}{y^3}\,q_7(x,y)\,{\rm d}T+\frac{y^4}{y_+^4}\,q_8(x,y)\,{\rm d}Z\right]\wedge{\rm Vol}_{S^2}\,.
 \label{eq:conformalC3}
\end{equation}
Here $x$ and $y$ are coordinates on the orbit space, unrelated to the axion variable \eqref{eq:defx} and to $y=e^{3v}$ in the main text. Thus the problem reduces to determining the eight functions $q_I(x,y)$, $I=1,\ldots,8$, on the two-dimensional domain $(x,y)\in[0,1]\times[0,1]$.

The particular form of the $(x,y)$ sector in Eq.~\eqref{eq:11d_ansatz} is a conformal gauge choice.  Indeed, the two-dimensional metric orthogonal to the Killing and spherical directions contains a single undetermined conformal factor, $q_2$, and no ${\rm d}x\,{\rm d}y$ component.  This completely fixes the local reparametrisation freedom of the orbit space, up to the usual residual global freedom fixed by the location of its boundaries.  In this gauge the Einstein equations can be divided into a set of second-order equations to be solved on the numerical grid and a pair of constraint equations.  The principal part of the former is governed by the inverse metric on the two-dimensional orbit space and is therefore elliptic in the exterior region.  The contracted Bianchi identities, together with the equation of motion for the three-form, imply first-order relations between the two remaining Einstein constraints.  With the appropriate boundary conditions, it is consequently sufficient to solve the elliptic subsystem: the constraints then vanish throughout the integration domain.  This is the higher-dimensional analogue of the conformal-coordinate construction described in Ref.~\cite{Horowitz:2024kcx}.

As a useful check, the ansatz contains the eleven-dimensional uplift of the non-extremal D0-brane as a special case.  Setting
\begin{equation}
 q_1=q_2=q_3=q_4=q_5=q_6=1,
 \qquad
 q_7=q_8=0,
\label{eq:D0_qs}
\end{equation}
restores the full $SO(9)$ symmetry and sets the four-form field strength to zero.  In particular,
\begin{equation}
 \frac{4\,{\rm d}x^2}{2-x^2}+x^2(2-x^2)\,{\rm d}\Omega_2^2+(1-x^2)^2\,{\rm d}\Omega_5^2={\rm d}\Omega_8^2 ,
\end{equation}
so that the transverse part of the metric becomes manifestly round. Introducing the standard D0-brane radial coordinate $r$ through
\begin{equation}
        y=\frac{r_0}{r},
        \qquad r_0=L\,y_+ ,
\label{eq:r_y_map}
\end{equation}
brings the metric to the standard eleven-dimensional uplift of the non-extremal D0-brane.  Thus $y=1$ is the event horizon, $r=r_0$, whereas the asymptotic region lies at $y=0$, corresponding to $r\rightarrow\infty$.

\subsection{Boundary conditions}

For the numerical implementation it is convenient to factor out part of
the known asymptotic behaviour.  We therefore introduce functions
$t_I(x,y)$ through
\begin{align}
 q_1&=1+y^2t_1,                 &
 q_2&=t_2,                       &
 q_3&=1+y^2t_3,                 \nonumber\\
 q_4&=t_4,                       &
 q_5&=t_5,                       &
 q_6&=1+y^2t_1+y^5t_6,        \label{eq:tvariables}  
 \\
 q_7&=H_1(x)t_7,                 &
 q_8&=H_1(x)t_8,                 &
 H_1(x)&=x^3(2-x^2)^{3/2}. \nonumber
\end{align}
The angular function $H_1$ is the $l=1$ harmonic appearing in the asymptotic analysis of Ref.~\cite{Costa:2014wya}.  The parametrisation \eqref{eq:tvariables} incorporates several properties of the asymptotic solution directly into the numerical variables: in particular, $q_1,q_3\to1$, $q_6-q_1=O(y^5)$, while the angular dependence of the non-normalisable mode of the three-form is already factored out.

The coefficient used in the conformal boundary conditions is dimensionless. We denote it by $\nu$. Near the boundary, $t_2=t_4=t_5=\eta_0^4$ and the physical coordinates are $t=L\eta_0^{-7}T$ and $u=L y_+\eta_0^2/y$ to leading order. The $q_7$ term in \eqref{eq:conformalC3} then gives a plane-wave flux of magnitude $3\nu/L$. Matching \eqref{eq:BMNplanewaveflux} and using \eqref{eq:massuplift} gives
\begin{equation}
 \nu\equiv\frac{L\mu}{3}=\frac{4m_\phi}{3L^6}.
 \label{eq:conformalmass}
\end{equation}
Thus $\nu$ is the conformal numerical source, while $\mu$ retains the BMN normalisation of \eqref{eq:BMNaction}. At the asymptotic boundary $y=0$, the independent boundary conditions can be written as
\begin{align}
 \partial_y t_1&=0,
&
 t_2^{5/2}t_6&=-\frac{\nu^2}{5y_+^5},
&
 \partial_y t_3&=0,
&
 \partial_y t_4&=0,
\nonumber\\
 \partial_y t_5&=0,
&
 \partial_y t_6&=0,
&
 t_2^{1/4}t_7&=\nu,
&
 t_2^{1/4}t_8&=\frac32\nu\,.
\label{eq:BCinfinity}
\end{align}

These conditions are equivalent to the boundary conditions used in Ref.~\cite{Costa:2014wya}, after accounting for the different gauge and choice of numerical variables.  In particular, the last two conditions turn on precisely the $\ell=1$ non-normalisable mode of the three-form, with the relative coefficient $3/2$ required by the eleven-dimensional field equations.  The remaining non-normalisable modes are absent. Notice that no independent condition on $\partial_y t_2$ is required. The asymptotic field equations themselves imply $\partial_y t_2|_{y=0}=0$ and
\begin{equation}
 t_2(x,0)=t_4(x,0)=t_5(x,0)=\eta_0^4 ,
\end{equation}
with $\eta_0$ independent of $x$.

At $y=1$ we demand regularity at the future event horizon. The horizon constraint implies
\begin{equation}
 \partial_x\left.\left(\frac{q_1}{q_2}\right)\right|_{y=1}=0,
\end{equation}
and hence
\begin{equation}
 q_1(x,1)=C_0\,q_2(x,1),
 \label{eq:horizonq1q2}
\end{equation}
where $C_0$ is a constant. This is the analogue of the condition $A=B$ imposed in Ref.~\cite{Costa:2014wya}: constancy of $q_1/q_2$ ensures that the surface gravity is uniform along the horizon. The constant $C_0$ is pure gauge and can be fixed using the residual scaling symmetry; we choose this freedom so that $C_0=1$. The remaining seven boundary conditions are obtained by expanding the dynamical equations about $y=1$ and solving them algebraically for the first radial derivatives $\partial_y t_I|_{y=1}$.  They are therefore Robin conditions of the form
\begin{equation}
 {\cal R}_I\!\left(
 t_J,\partial_x t_J,\partial_x^2t_J,\partial_y t_J
 \right)\bigg|_{y=1}=0,
 \qquad I=2,\ldots,8.
\label{eq:horizonRobin}
\end{equation}
Their explicit form is lengthy and will not be reproduced here.

The appearance of Robin rather than Neumann conditions is merely a consequence of our radial coordinate.  A smooth radial coordinate $\rho$ at the bifurcation surface satisfies $\rho^2\propto1-y$. Smoothness requires the fields to be even functions of $\rho$, but this allows a generic expansion linear in $1-y$ and hence a non-vanishing $\partial_y t_I|_{y=1}$.  This is equivalent to the horizon regularity conditions used in Ref.~\cite{Costa:2014wya}, where a coordinate proportional to $\sqrt{1-y}$ was introduced explicitly.

At the two symmetry axes the conditions follow directly from smoothness. At $x=0$ the $S^2$ shrinks to zero size, and absence of a conical singularity requires
\begin{equation}
 t_2(0,y)=t_4(0,y),\qquad  \left.\partial_x t_I\right|_{x=0}=0,  \qquad I=1,3,4,5,6,7,8 .
\label{eq:BCx0}
\end{equation}
Similarly, at $x=1$ the $S^5$ shrinks and we impose
\begin{equation}
 t_2(1,y)=t_5(1,y),\qquad
 \left.\partial_x t_I\right|_{x=1}=0,
 \qquad I=1,3,4,5,6,7,8 .
\label{eq:BCx1}
\end{equation}
The field equations then imply the corresponding Neumann condition for $t_2$.  These are precisely the regularity conditions of Ref.~\cite{Costa:2014wya}, expressed in conformal gauge.

We compute the energy of these solutions directly in eleven dimensions, following the covariant Noether-charge construction of Ref.~\cite{Dias:2019Noether}. We first use coordinates in which the asymptotic time and circle have fixed normalisations, and vary both $y_+$ and the deformation parameter $\nu$. We evaluate the charge variation for the corresponding Killing fields, including the metric, four-form and Chern--Simons contributions, and integrate it over the $S^8\times S^1$ at the boundary. At the horizon the same variation gives $T\,\delta S$. The boundary result and the scaling relation then determine the energy and mass response up to a source-dependent integration term, which we fix by requiring the supersymmetric branch to have zero energy. This gives an eleven-dimensional determination of the energy independently of the reduced calculation in section~\ref{sec:holoenergy}.

\begin{figure}[t]
\centering
\includegraphics[width=\textwidth]{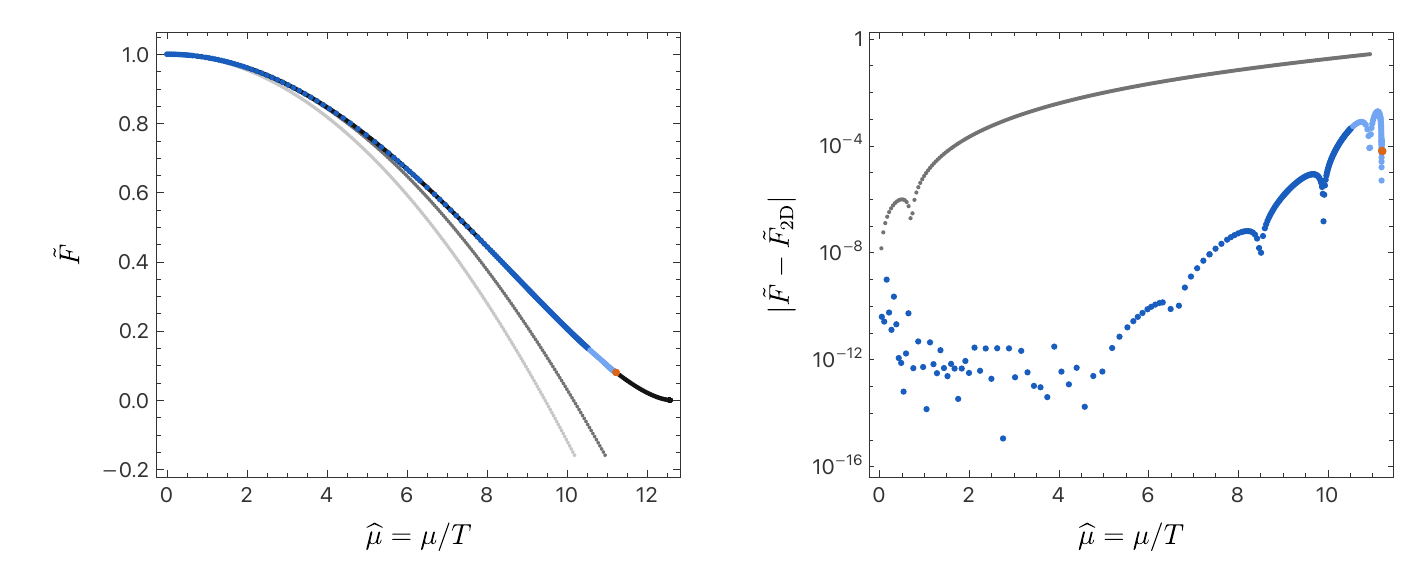}
\caption{Free-energy comparison in a common BMN mass convention. The light grey points are the saved 2014 data of Ref.~\cite{Costa:2014wya} with the conversion \eqref{eq:CGPSmassconversion}, and the dark grey points are the same data with the rescaled axis \eqref{eq:CGPSrescaling}. The black points are the two-dimensional result, and the dark blue points are the $25\times45$ conformal-coordinate result. The light blue and orange points extend the conformal branch by arclength continuation on $25\times45$ and $31\times55$ grids, respectively. The right panel shows the absolute differences from the two-dimensional free energy, evaluated with a fifth-order interpolant, on a logarithmic scale. The growth of the $25\times45$ residual beyond $\widehat\mu\simeq10$ reflects the finite resolution of that grid; the better-resolved $31\times55$ points agree with the two-dimensional result to $6\times10^{-5}$.}
\label{fig:conformalcomparison}
\end{figure}

Figure~\ref{fig:conformalcomparison} compares the resulting free energy with the two-dimensional calculation and with the curve displayed in Ref.~\cite{Costa:2014wya}. We have converted all three horizontal axes to $\widehat\mu=\mu/T$: the saved two-dimensional data use $m_\phi/T=\widehat\mu/4$, whereas the conformal data file uses $\nu/T=\widehat\mu/3$ in units $L=1$. The older paper defines $\widehat\mu_{\rm CGPS}=7\widehat\mu/(12\pi)$, as in \eqref{eq:CGPSmassconversion}. Plotted in this way, however, its free-energy curve disagrees with ours already at small $\widehat\mu$, where both calculations should be accurate; at $\widehat\mu\simeq0.9$ the difference is $1.2\times10^{-3}$. This disagreement is almost entirely removed by a rescaling of the horizontal axis. Fitting $\widehat\mu=k\,\widehat\mu_{\rm CGPS}$ to the two-dimensional curve over $0<\widehat\mu_{\rm CGPS}<x$ gives $k=5.7888$, $5.7900$, $5.7919$ and $5.7971$ for $x=0.1$, $0.15$, $0.2$ and $0.3$, and matching the coefficients of the $\widehat\mu^2$ terms gives $k=5.788$. We therefore compare with the 2014 data using
\begin{equation}
 \widehat\mu=k\,\widehat\mu_{\rm CGPS},\qquad k=5.79\pm0.01,
 \label{eq:CGPSrescaling}
\end{equation}
in place of $\widehat\mu=(12\pi/7)\widehat\mu_{\rm CGPS}\simeq5.386\,\widehat\mu_{\rm CGPS}$. With this rescaling the 2014 curve agrees with ours to $4\times10^{-6}$ at $\widehat\mu=0.9$, and departs from it by $10^{-3}$ only beyond $\widehat\mu\simeq2.9$ and by $10^{-2}$ beyond $\widehat\mu\simeq5.1$. The value of $k$ agrees to $0.02\%$ with $(12\pi/7)(4/3)^{1/4}$, which suggests a specific normalisation error in the horizontal axis of the 2014 plot, but we have not been able to trace its origin. The zero of the rescaled curve lies at $\widehat\mu\simeq10.15$, rather than at the value $\widehat\mu\simeq9.44$ obtained with \eqref{eq:CGPSmassconversion}. Reference~\cite{Costa:2014wya} derives \eqref{eq:CGPSmassconversion} explicitly and uses it to quote its critical temperature. Until the origin of this axis mismatch is identified, the rescaling should be regarded as an empirical alignment of the plotted curve, not as a correction of that critical temperature.

The location of the zero of the 2014 curve is, moreover, very sensitive to the procedure used to obtain it. Ref.~\cite{Costa:2014wya} fitted $\tilde S$ by a polynomial in $\widehat\mu_{\rm CGPS}$ including odd powers and integrated \eqref{eq:firstlaw} term by term, which multiplies the coefficient of $\widehat\mu^n$ by $14/(14-5n)$. For $n=3$ this factor is $-14$, since the homogeneous solution of \eqref{eq:firstlaw} behaves as $\widehat\mu^{14/5}$. Since $\tilde S$ is even in $\mu$, odd coefficients are artefacts of the fit, and they are strongly amplified. Applying the same procedure to our entropy data over the same range, $\widehat\mu_{\rm CGPS}\le1.76$, gives zeros between $T/\mu\simeq0.086$ and $0.142$, depending only on the degree of the fit. Restricting to even powers gives $T/\mu\simeq0.080$--$0.082$, consistent with the end point $1/(4\pi)\simeq0.0796$. We therefore attribute the crossing found in Ref.~\cite{Costa:2014wya} to this step rather than to the solutions themselves.

The close agreement of the new conformal data with the two-dimensional curve extends through $\widehat\mu\simeq10.5$, well beyond the point where the 2014 curve departs from ours. The absolute residual in the right panel also makes the finite resolution of the $25\times45$ conformal run visible near the end of its range. An arclength continuation in the same formulation extends the conformal branch, as shown by the light blue and orange points in Fig.~\ref{fig:conformalcomparison}, to $\widehat\mu\simeq11.20$ on the $25\times45$ grid and $\widehat\mu\simeq11.22$ on a $31\times55$ grid. On the $31\times55$ grid, whose Chebyshev tails remain below $6\times10^{-7}$, the conformal free energy agrees with the two-dimensional one to $6\times10^{-5}$ at $\widehat\mu=11.22$. On the $25\times45$ grid the tails grow to $2\times10^{-5}$ and the difference to $2\times10^{-3}$ near $\widehat\mu\simeq11.1$, reflecting its finite resolution there.

\bibliographystyle{JHEP}
\bibliography{biblio}

\end{document}